%% file: pinning_flat.tex
\documentclass[aip,jcp,reprint ,superscriptaddress]{revtex4-1}
\usepackage[dvips]{graphicx,rotating}
\usepackage{amssymb}
\usepackage{amsmath}
\usepackage{latexsym}
\usepackage{epsfig}
\usepackage{bm}
\usepackage{psfrag}
\usepackage{url}
\usepackage{etoolbox}
\usepackage{amsfonts,float}
\usepackage{gensymb}
\usepackage{caption}
\usepackage{subcaption}
\usepackage{xspace}

\usepackage{graphicx}
\usepackage[usenames,dvipsnames]{color}
\usepackage{rotating}

\newcommand{\Sconf}{S_{\rm MS}}
\newcommand{\sconf}{s_{\rm MS}}

   \newcommand{\red}[1]{{#1}}

\newcommand{\rlj}[1]{{#1}}

\begin{document}


\title{Effects of random pinning on the potential energy landscape of a supercooled liquid}

\author{S. P. Niblett}
\email[]{sn402@cam.ac.uk}
\affiliation{University Chemical Laboratories, Lensfield Road, Cambridge CB2 1EW, UK}
\author{V. K. de Souza}
\email[]{vkd21@cam.ac.uk}
\affiliation{University Chemical Laboratories, Lensfield Road, Cambridge CB2 1EW, UK}
\author{R. L. Jack}
\affiliation{University Chemical Laboratories, Lensfield Road, Cambridge CB2 1EW, UK}
\affiliation{Department of Applied Mathematics and Theoretical Physics, University of Cambridge, Wilberforce Road, Cambridge, Cambridge CB3 0WA, UK}
\author{D. J. Wales}
\email[]{dw34@cam.ac.uk}
\affiliation{University Chemical Laboratories, Lensfield Road, Cambridge CB2 1EW, UK}


\begin{abstract}
We use energy landscape methods to investigate the response of a supercooled
liquid to random pinning.  We classify the structural similarity of different
energy minima, using a measure of overlap.
This analysis reveals a correspondence between distinct particle packings (which are
characterised via the overlap) and funnels on the energy landscape (which are
characterised via disconnectivity graphs).  As the number of pinned particles
is increased, we find a crossover from a fluid state at low pinning to a glassy
state at high pinning, in which all thermally accessible minima are
structurally similar to each other.  We discuss the consequences of these
results for theories of randomly pinned liquids.  We also investigate how the
energy landscape depends on the fraction of pinned particles, including the
degree of frustration and the evolution of distinct packings as the number of
pinned particles is reduced.

\end{abstract}


\maketitle



\section{Introduction}
\label{sec:introduction}

A structural glass is a material that is mechanically solid but has an
amorphous, liquid-like microstructure.\cite{BerthierB11} Glasses are normally
produced by rapidly supercooling a liquid, which causes the constituent
particles to move increasingly slowly until the system becomes solid on the
experimental time scale.  Several theories aim to describe this dynamical
slowing down~\cite{LubchenkoW07,ChandlerG10,gotzes88} by proposing that cooling
the system brings it close to a critical point at which structural relaxation
stops completely.
This critical point might be thermodynamic in
origin,~\cite{adamg65,LubchenkoW07} or a purely dynamical effect, and the
associated singularity might occur at some finite
temperature,~\cite{LubchenkoW07,gotzes88} or in a limit where the temperature
tends to zero.~\cite{ChandlerG10}

Cammarota and Biroli~\cite{CammarotaB11} proposed that by gradually
\emph{pinning} (immobilising) some of the particles in a supercooled liquid,
one might observe a singularity at which structural relaxation of the remaining
(unpinned) particles stops completely. This has been termed the
\emph{random-pinning glass transition} (RPGT). The associated singularity takes
place at finite temperature, and shares many features with the singularity that
occurs in random first-order transition (RFOT) theory.~\cite{LubchenkoW07}  
In both the RPGT and RFOT transitions, slow (glassy) dynamics originates in a
reduction of the entropy: a supercooled liquid may adopt many different
amorphous structures, but a glass is a system that is localised on the observation
time scale into a single metastable state, corresponding to a group of potential 
energy minima.

The theoretical proposal of Ref.~\onlinecite{CammarotaB11} has its roots in mean-field theory, and it is not clear whether these predictions are valid in physical (three-dimensional) fluids.  Several numerical studies have investigated the effects of random pinning,~\cite{ScheidlerKBP02,KimMS11,BerthierK12,KobB13, SzamelF13,FullertonJ14,OzawaKIM15,CammarotaS16,ChakrabartyDKD16} but 
have not yet established (or ruled out) the existence of an RPGT.  Performing this test and accurately characterising the RPGT would require a comprehensive finite-size scaling analysis. However, the corresponding calculations converge slowly and the size-scaling approach is currently very expensive. 

In this work, we use geometry optimisation methods~\cite{Wales2003} to explore
 the potential energy landscape (PEL) of a randomly-pinned glassy fluid. Analysis of the
PELs of glasses has a long
history.\cite{Goldstein69,StillingerW82,StillingerW84,SastryDS98,SchroderSDG00,DoliwaH03,
DoliwaH03b, Heuer08,deSouzaW08,NiblettDSW16}
A key advantage of using this approach to study the RPGT is that
geometry optimisation methods such as basin-hopping (BH) global
optimisation\cite{lis87,walesd97a} 
and discrete path sampling (DPS)\cite{Wales02,Wales04} can treat activated relaxation events with arbitrarily high energy barriers.
Exploring the accessible configuration space via DPS is efficient
compared with conventional methods such as molecular dynamics that require
waiting for many such events, which become increasingly rare as the glass
transition is approached.


Our study combines energy landscape methods with an idea from mean-field
theory, that a useful order parameter is the overlap $Q$, which measures the
structural similarity of two configurations for a glassy
system.~\cite{FranzP97,MezardP99}  We classify minima of the PEL according to
their overlap, so that the RPGT (if it exists) is associated with localisation
of the system, on long time scales, in a region of the landscape where all configurations are
structurally similar (high-overlap).~\cite{CammarotaB11}  Our results are
restricted to small systems ($N=256$ particles), but we do find a crossover on
increasing the number of pinned particles, from  low-overlap to high-overlap.
These results are consistent with the predictions of
Ref.~\onlinecite{CammarotaB11}, although the restriction to small systems means
that we cannot distinguish whether the system has a smooth crossover from low-
to high-overlap, or whether there might be a true phase transition.

In addition, we investigate how the PEL changes as particles are \red{unpinned.}  We
discuss the relationship between funnels on the energy landscape, the packing
of the particles in space, and the overlap $Q$.  

The structure of the paper is as follows: Sec.~\ref{sec:Methods} describes our
model and methods; and Sec.~\ref{sec:minimaresults} characterises the crossover
from low- to high-overlap, including a summary of the implications for the RPGT
in Sec.~\ref{sec:xover-discuss}.  Then Sec.~\ref{sec:frustration} analyses the
varying degree of frustration of the landscape, following
Ref.~\onlinecite{deSouzaSNFW17}.  In Sec.~\ref{sec:LandscapeEvolution} we
analyse in more detail the dependence of the energy landscape on the number of
pinned particles, by following the behaviour of distinct particle packings as
the number of pinned particles is reduced.  Finally, Sec.~\ref{sec:Conclusions}
summarises our conclusions.

\section{\label{sec:Methods}Methods}

\subsection{\label{sec:Model}Model}

We consider a binary Lennard-Jones (BLJ) fluid. There are two atom types,
larger A atoms and smaller B atoms, interacting via Lennard-Jones potentials
with the popular Kob-Andersen parameter set.\cite{KobA95a} This choice allows
comparison with earlier numerical studies of the
RPGT,\cite{BerthierK12,OzawaKIM15,ChakrabartyKD15} and with earlier work on the
energy landscapes of glass-forming
liquids.\cite{deSouzaW05,deSouzaW06,deSouzaW06b,deSouzaW08,deSouzaW09,NiblettBWD17}

\rlj{The truncation of the interaction potential uses the Stoddard-Ford quadratic shifting scheme, \cite{StoddardF73} which ensures that the potential and its gradient are both continuous, as required for landscape analysis.  Full details of the potential are given in Ref.~\onlinecite{deSouzaW05}: the truncation range is  $r_c = 2.5\,\sigma_{\rm AA}$.}
The system size is $N=256$ atoms, and we use a periodically-repeated cell with
fixed number density 1.2$\,\sigma_{\rm AA}^{-3}$.  
$\mathbf{X}=(\bm{r}_1,\bm{r}_2,\dots,\bm{r}_N)$ will denote a vector containing the
positions of all particles.

Following convention, all lengths, energies, temperatures and times are given
in reduced units, which can be expressed in terms of $\sigma_{\rm AA}$,
$\epsilon_{\rm AA}$ and $m$, the mass of an A or B atom.


\subsection{\label{sec:PinningMethod}Pinning Particles}

The random pinning method~\cite{CammarotaB11,BerthierK12,OzawaKIM15} makes use
of a \emph{reference configuration} $\mathbf{X}^*$ that is sampled from the
equilibrium (Boltzmann) distribution at some temperature $T_0$.
Reference configurations were taken from molecular dynamics simulations, 
which were confirmed to be locally ergodic using the Mountain-Thirumalai fluctuation metric.\cite{ThirumalaiMK89,ThirumalaiM93,deSouzaW05}

Let $c$ be the fraction of pinned particles, which means that $M=\left \lfloor{cN}\right \rfloor$ 
particles are chosen independently at random. The
notation $\lfloor x \rfloor$ indicates the largest integer that is less than or
equal to $x$.  The positions of these $M$ particles are fixed, and one
considers the energy of the system as a function of the position of the
remaining (unpinned) particles. The potential energy surface (i.e.~PEL) and
associated Boltzmann distribution depend on $\mathbf{X}^\ast$ and on the set of
pinned particles.
Following the literature on spin glasses and other disordered systems, we refer to the combination of the reference configuration and the 
set of pinned particles as a \emph{realisation of the disorder}.  To obtain robust results one should average over many such realisations.

For each realisation of the disorder and each configuration $\mathbf{X}$ of the
unpinned particles, the LBFGS algorithm\cite{Nocedal80,LiuN89} was used to
minimise the energy (always with the pinned particles fixed), so that each
configuration can be associated to a local minimum of the PEL.   The local
minimum associated with the reference configuration $\mathbf{X}^*$ is the
\emph{reference minimum} $\mathbf{X}_0$.

\red{All calculations performed in this paper, except those in sec.~\ref{sec:lowT}, used reference temperature $T_0 = 0.5\,\epsilon_{\rm AA}/k_{\rm B}$. The results of Ozawa {\em et al.}\cite{OzawaKIM15} suggest that this should be the highest reference temperature at which the RPGT transition would be clearly observed for our model.}

\subsection{\label{sec:Overlap}Comparing Structures: the overlap}

Mean-field theory proposes that a useful order parameter in glassy systems is 
the overlap between  two configurations $\mathbf{X}_i,\mathbf{X}_j$.
\rlj{Several definitions are possible for the overlap,\cite{BiroliBCGV08,Berthier13,OzawaKIM15,DasCK17} and the general expectation\cite{BiroliC17} is that all definitions should lead to similar behaviour as long as two configurations have high overlap if and only if they are structurally similar.}
%
%
Following Ref.~\onlinecite{FullertonJ14}, our pinning procedure is independent of the particle types but 
the overlap depends only on the type-A particles:
\begin{equation}
  Q(\mathbf{X}_i,\mathbf{X}_j) = \frac{1}{N_m}\sum_{k=1}^{N_m} \Theta(a - |\mathbf{r}_{k,i} - \mathbf{r}_{k,j}|).
\end{equation}
Here, $k$ runs over the set of unpinned A-type atoms, and $N_m$ is the number
of such atoms. $\Theta$ is the Heaviside step function, and $\mathbf{r}_{k,i}$
is the position vector of atom $k$ in configuration $\mathbf{X}_i$. Also, $a$
is a length scale parameter, which we set to $0.3\,\sigma_{\rm AA}$ following
earlier work.\cite{OzawaKIM15,KobC14,Berthier13} Before calculating the overlap, 
permutational alignment is performed for $\mathbf{X}_i$ and $\mathbf{X}_j$ 
to account for indistinguishability of the mobile A atoms,
using a shortest augmenting path algorithm.\cite{JonkerV87,WalesC12} 
\rlj{This ensures that (for example) swapping the positions of two particles does not affect the overlap.}


If $\mathbf{X}_i$ and $\mathbf{X}_j$ are very similar then
$Q(\mathbf{X}_i,\mathbf{X}_j)\approx 1$, which is the largest possible value.
The smallest possible value is $Q= 0$, and independent random configurations
typically have small values  $Q\approx 0$.  Based on the decay of overlap that
happens during $\beta$-relaxation,~\cite{BerthierBBKMR07} and on other previous
work,~\cite{BerthierJ15,OzawaKIM15} we introduce a threshold parameter
$Q^*=0.7$ such that if $Q(\mathbf{X}_i,\mathbf{X}_j)>Q^*$ then we identify
$\mathbf{X}_i$ and $\mathbf{X}_j$ as being structurally similar.

The overlap of an arbitrary configuration $\mathbf{X}$ with the reference minimum $\mathbf{X}_0$ is
\begin{equation}
Q_0(\mathbf{X}) = Q(\mathbf{X},\mathbf{X}_0) .
\end{equation}

%

\begin{figure}
  \includegraphics[width=\columnwidth]{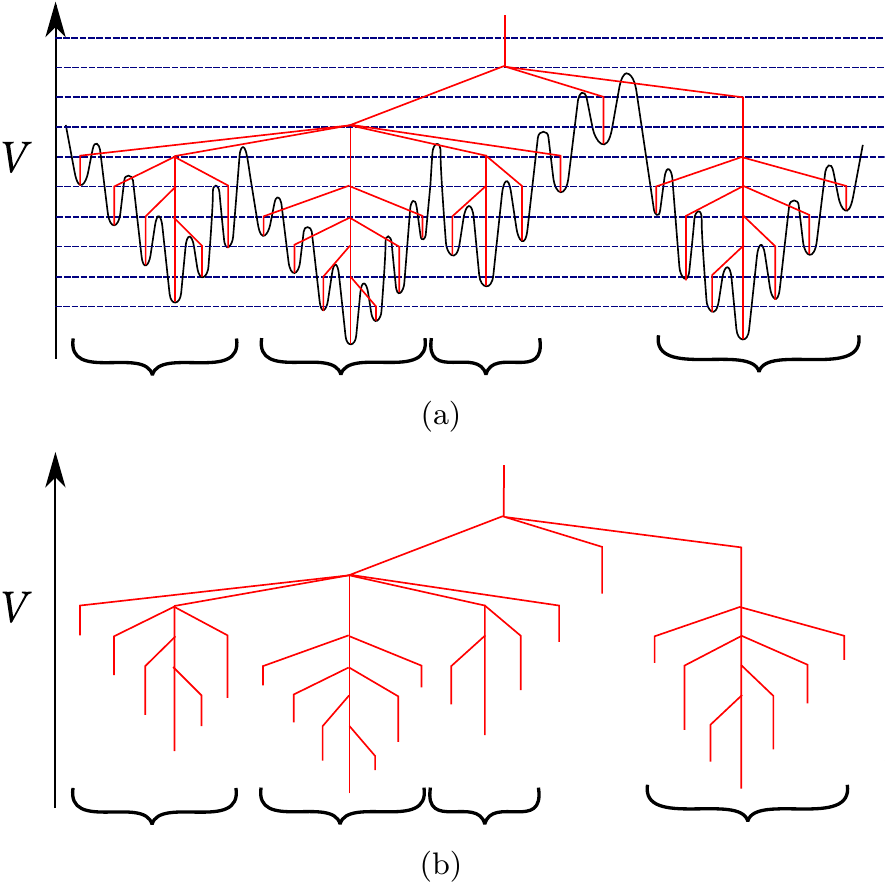}
  \caption{\label{fig:landscapecartoons} Top: Cartoon glassy PEL (in black)
showing the disconnectivity graph construction (in red). Dashed blue lines indicate
energy thresholds, brackets show the approximate extent of local funnels. Bottom:
the disconnectivity graph that results from this cartoon landscape.}
\end{figure}

\subsection{\label{PELMethods}Potential Energy Landscape Methodology}



The PEL of a pinned system is given by $V(\mathbf{X})$, which is a
3\,$N_m$-dimensional function because the positions of the pinned atoms are
constant. In the following we adapt standard geometry optimisation methods to
identify local minima of $V(\mathbf{X})$, which control the thermodynamics at
low temperatures, and transition states (stationary points with Hessian index
one) which control the dynamics.

We classify minima of the pinned landscape according to their energies $V$ and
their overlaps $Q_0$ with the reference minimum. We define a potential energy density of minima
$g_{\rm IS}$ such that $g_{\rm IS}(V,Q_0)\, dV \, dQ$ is the number of minima
with energies between $V$ and $V+dV$ and overlaps between $Q_0$ and $Q_0+dQ$.
We can also define a landscape entropy, $S_{\rm IS}(V) = k_\text{B}\ln g_{\rm IS}(V)$,\cite{SciortinoKT99,Wales2003}
\red{where $g_{\rm IS}(V)\,dV$ is the number of minima with energies between $V$ and $V+dV$.}
The subscript IS refers to ``inherent structures'', an alternative nomenclature for PEL local minima.\cite{StillingerW82}

\subsubsection{\label{sec:BHPT}Exploring the PEL: Identifying Minima}

To locate minima of the PEL, we use the basin-hopping\cite{lis87,walesd97a} and basin-hopping parallel tempering\cite{StrodelLWW10} methods.
The \emph{basin-hopping} algorithm begins from a local minimum \red{(see secs.~\ref{sec:ReturnTrajs} and \ref{sec:MinimaHistograms} for different approaches to choosing the initial minimum)}. In each step, a
new local minimum is proposed by perturbing the structure and performing local
energy minimisation. The step is accepted or rejected using a Metropolis
criterion \red{with effective temperature $T_{\rm BH}$}. If the step is accepted, the original minimum is stored in a database and the
next step begins from the new minimum. Otherwise, the algorithm returns to the
original minimum for the next step.

The two main parameters in this procedure are \red{the maximum size of the structural perturbation, and the effective temperature $T_{\rm BH}$. These parameters control the rates at which new minima are discovered and accepted, respectively, but the true PEL being sampled is independent of both. 
The parameters are usually selected through trial and error to achieve a target acceptance rate for new minima, and may be varied dynamically during the calculation to assist this objective.} 

Basin-hopping locates low-energy minima efficiently, because the energy
minimisation step removes downhill potential energy barriers between adjacent
local minima. However, when a PEL contains many \emph{funnels} (see below) the
algorithm may become temporarily trapped in a single low-energy region and not explore
alternative competing structures. \red{Using a low $T_{\rm BH}$ exacerbates this problem, but using a high $T_{\rm BH}$ means that the low-energy regions may not be sampled adequately.}

To avoid this problem, \emph{basin-hopping parallel tempering} (BHPT)
can be applied.\cite{StrodelLWW10} In BHPT, one considers $N_r$ replicas of the
basin-hopping algorithm, each with a different value of $T_{\rm BH}$.  By
exchanging different configurations between the replicas, high-energy regions
of the landscape may be crossed more efficiently.


Basin-hopping and BHPT both explore local minima efficiently, and can (in
principle) be used to identify all minima on a PEL.  
Our strategy here is to generate a histogram of minimum energies and overlaps,
which we denote by $\rho_{\rm IS}(V,Q_0)$.  In the limit where the algorithm
samples the PEL exhaustively for a given range of $V$ and $Q_0$
\begin{equation}
\rho_{\rm IS}(V,Q_0) \to g_{\rm IS}(V,Q_0).
\label{equ:rho-g}
\end{equation}
In practice, the landscape is not explored exhaustively so there are systematic differences between $\rho_{\rm IS}$ and $g_{\rm IS}$, but we expect the important qualitative features of $g_{\rm IS}$ to be mirrored by $\rho_{\rm IS}$. This point is discussed in more detail during the analysis of the results.

\subsubsection{\label{sec:TSsearches}Transition State Searches}

Many properties of physical systems can be predicted by considering the
statistics of local minima on the PEL, together with the transition states 
that connect them.
To obtain information about transition states, we start from a database of
minima and use double-ended searches, which take two local minima as input and
identify a discrete path between them. A discrete path is a sequence of
transition states and intermediate minima connected by steepest-descent
paths.\cite{Wales02,Wales04}. Specifically, the
doubly-nudged\cite{TrygubenkoW04,SheppardTH08} elastic
band\cite{HenkelmanUJ00,HenkelmanJ00} (DNEB) algorithm is used to construct an
approximate minimum energy pathway between pairs of minima. Structures
corresponding to the local maxima on this pathway are candidate transition
states, which are refined accurately using hybrid eigenvector-following
(HEF).\cite{munrow99,Wales2003,ZengXH14}  If a complete pathway 
between the original pair of minima has not been identified after one DNEB/HEF cycle,
the Dijkstra algorithm\cite{Dijkstra59} is used with
a suitable distance metric\cite{CarrTW05} to select another pair of minima and
another cycle is attempted. This procedure is repeated until the
original pair of minima have been connected. Double-ended searches often add
intermediate minima to the database as well as the connecting transition
states. 

Once transition states are known, energy barriers between minima may be
measured. The overall barrier height from minimum $A$ to $B$ is defined as the energy
difference between $A$ and the highest transition state on the minimum-energy
pathway from $A$ to $B$.

\subsubsection{\label{sec:DGraphMethods}Disconnectivity Graphs}

After many transition state searches, one obtains a database of energy minima connected by transition states.
To analyse the landscape, it is useful
to generate a disconnectivity graph,\cite{BeckerK97,Wales2003} which can be
interpreted by visual inspection.  In these graphs, energy minima are
represented as points, whose heights indicate the corresponding potential
energy.  These points are connected (upwards) to branching points: the energy
of a branching point is (close to) the energy of the transition state that
connects the minima below it. Transition state energies are rounded up to discrete
energy levels to produce a clear visual representation. 
The process of generating a disconnectivity graph from a model glassy landscape is shown schematically in fig.~\ref{fig:landscapecartoons}.

Disconnectivity graphs\cite{BeckerK97,Wales2003} faithfully represent the energies of the local minima and the energy barriers between them, as determined from the transition states.

\subsection{Landscape features: Funnels, metabasins, packings, and the configurational entropy}
\label{sec:funnels-etc}

A central motivation for random pinning studies~\cite{CammarotaB11,KobB13} is
that they enable (in principle) the \rlj{accessible configuration space} of a system
to be
varied, without changing the temperature or the liquid structure, and hence
without requiring extensive equilibration at low temperatures.  Within
mean-field theory, \rlj{this effect is controlled by the 
\emph{configurational entropy}, which measures the equilibrium occupation probabilities of metastable states.
(Note that this quantity is not an experimental observable, and must be distinguished from the configurational part of the total entropy,
which is defined in terms of the potential energy density of states.\cite{Wales2003})}
Within mean-field theory, the \rlj{configurational entropy} vanishes continuously at some finite pinning fraction $c=c^*$; this
is the RPGT.  

By investigating how the PEL changes with pinning, we can test
how this mean-field prediction plays out in the pinned BLJ liquid.  
\rlj{However, while the configurational entropy is a well-defined quantity in mean-field models, it does not have a unique definition in 
finite-dimensional systems} \red{where interactions have finite range,} \rlj{since metastable states have finite lifetimes in this case. 
Nevertheless, metastable states}
can be identified as regions of configuration space within which
the system remains (dynamically) localised over sufficiently long time periods.
See Ref.~\onlinecite{BiroliK01} for a discussion of how this construction can be
applied consistently in both finite-dimensional and mean-field systems.

We consider three ways of identifying candidates \rlj{for such} states.  We claim in the following that all these definitions
identify a similar set of candidate metastable states.

A \emph{local funnel} on the PEL  is a group of minima 
for which barriers to reach a lower-energy minimum (downhill barriers) are systematically smaller than barriers to reach a higher-energy minimum (uphill barriers). 
Funnels are usually identified informally by visual
inspection of disconnectivity graphs (see Fig.~\ref{fig:landscapecartoons}).
Glassy landscapes typically contain many local funnels in the same energy
range. Energy barriers between funnels are typically larger than barriers
within funnels.

A \emph{metabasin} is a group of minima between which dynamical transitions are
rapid and easily reversible.\cite{DoliwaH03,DoliwaH03b} Transitions between states 
in different metabasins are slower and should
correspond to structural relaxation of the glass-former.  In this sense,
metabasins are similar to metastable states (albeit with finite lifetimes).
Local funnels often correspond to metabasins.\cite{deSouzaW08}

A \emph{packing} of the particles is a group of minima with high mutual overlap and low overlap with other packings.  Within mean-field theories, this method can be used to identify  metastable states, but we emphasise that their definition does not include any dynamical information.

\rlj{With these definitions in hand, we can
define a measure of entropy that corresponds to the configurational entropy defined within
mean-field theory.  Suppose that for low temperatures $T$ the
thermally-accessible configuration space for pinning fraction $c$ can be broken up
into a set $\Lambda(c,T)$ of distinct packings, and that the canonical
(Boltzmann) equilibrium occupation probability for packing $i$ is $p_i(c,T)$.  Then a
metastable state entropy, $S_{\rm MS}$, can be defined as
\begin{equation}
S_{\rm MS}(c,T) = - \sum_{i \in \Lambda(c,T)} p_i(c,T) \log p_i(c,T)
\label{equ:Sc}
\end{equation}
which depends implicitly on the system size $N$.
To the extent that funnels and metabasins coincide with packings, the sum in (\ref{equ:Sc}) can be replaced by a sum over metabasins or funnels.  
On taking the thermodynamic limit, $S_{\rm MS}$ per particle is
\begin{equation}
\sconf(c,T) = \lim_{N\to\infty} \frac{\Sconf(c,T)}{N}.
\end{equation} 
The mean-field prediction is that $\sconf(c,T)$ vanishes continuously at the RPGT ($c=c^*$).
Hence, for $c<c^\ast$ mean-field theory predicts $s_{\rm MS} > 0$ but for
$c>c^\ast$ then $s_c=0$.  Recalling (\ref{equ:Sc}), this result means that for
finite
systems and $c>c^\ast$, the number of metastable states with significant occupation probabilities is
sub-exponential in $N$: we will typically find that a single metastable state
becomes dominant.
}

These hypotheses will be tested in the next Section.  \rlj{We do not attempt to measure $s_{\rm MS}$ directly,
but we do find that for small $c$ then many packings contribute to the sum in~(\ref{equ:Sc}), while for large $c$ then only a single
packing contributes.  These results show how the PEL evolves during this process, 
complementing previous studies\cite{KobB13,OzawaKIM15}, which observed a dramatic
decrease in the entropy of a supercooled liquid on increasing the
pinning fraction $c$ through a critical value $c^\ast$. }


\rlj{We note that $S_{\rm MS}$ has similarities with several other measures of entropy.}
The landscape
entropy $S_{\rm IS}$ is defined in a microcanonical framework
in terms of the potential energy density of minima.\cite{SciortinoKT99,Wales2003}  \rlj{This landscape entropy can be estimated as the difference 
between the total entropy of the system and its vibrational entropy.\cite{SciortinoKT99}}
$S_{\rm IS}$ is not the same as $S_{\rm MS}$ because \rlj{a metastable state does not correspond to a single energy minimum}:
the waiting time within minima is often quite short,\cite{Cavagna09} and $S_{\rm IS}$ is not expected to vanish at
the RPGT.  
\rlj{Other measures of metastable state entropy include the free energy cost required to localise the system
in a state of high-overlap,\cite{BerthierC14} and direct counting of structural motifs\cite{KurchanL11} -- our expectation is that these quantities should behave in a similar way to $S_{\rm MS}$.}

\section{\label{sec:minimaresults}Results: Potential Energy Minima and Landscape Organisation}

\subsection{\label{sec:ReturnTrajs}Return Times}

To illustrate that increasing $c$ dramatically reduces the number of states with
significant equilibrium occupation probability, 
we used basin-hopping to explore low-energy minima on the \red{landscapes corresponding
to several different pinning fractions}.  The reference
minimum $\mathbf{X}_0$ has a low energy, and we expect it to be part of an
accessible state for all $c$.  

Let $\mathbf{X}(s)$ be the PEL minimum obtained
after $s$ basin hopping steps, and let $Q_0(s) = Q(\mathbf{X}_0,\mathbf{X}(s))$
be the overlap of this minimum with the reference minimum.  \red{We extracted 
the initial minima $\mathbf{X}(0)$ from molecular dynamics simulations of the pinned
system with very high temperatures, so that $Q_0(0)$ would be small. 30 basin-hopping 
calculations were performed for each value of $c$, each with a structurally distinct
initial minimum.}

If the only accessible metastable state is the one containing the reference minimum,
then we expect \red{each basin-hopping calculation} to converge to that state:
$Q_0(s)>Q^\ast$ for large $s$.  On the other hand, if there are many accessible
states, one expects the algorithm to explore regions that are different from the
reference configuration, \red{so that} $Q_0(s)$ will remain small.

\begin{figure}
    \begin{center}
      \includegraphics[width=0.5\textwidth]{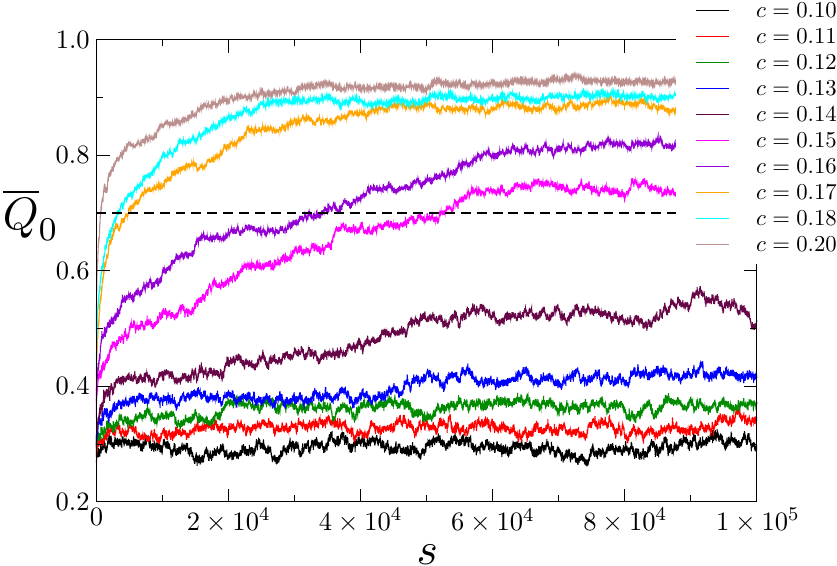}
    \end{center}
  \caption{\label{fig:returntrajs}Plots showing how basin-hopping calculations explore overlap space at several different pinning fractions. $s$ denotes the number of basin-hopping steps taken, and $\overline{Q}_0(s)$ is the average of the overlap $Q_0$ for 30 basin-hopping runs with independent starting points. The dashed horizontal line shows $Q_0=Q^\ast$, which is the threshold used to define when structures are similar to the reference.}
\end{figure}

Fig.~\ref{fig:returntrajs} shows results for a single reference configuration
($T_0=0.5$) and various pinning fractions.  We used an initial
temperature parameter $T_{\rm BH}=5$, which was adjusted during the sampling to
maintain an acceptance rate of 70\%. \red{Other parameter sets were found to give
similar results, but the calculations were less efficient.} 

\red{Fig.~\ref{fig:returntrajs} shows that calculations with $c\geq0.15$ tend to a
large-$s$ limit of $\overline{Q}_0(s)>0.7$,} indicating that the system \red{approaches} the
reference minimum.  For $c\leq0.13$ we find $\overline{Q}_0(s)\lesssim0.4$ \red{for all $s$,}
indicating that the system is exploring a larger region of configuration space.  
These observations suggest a rather sharp crossover in the overlap as $c$ is increased, 
consistent with the predictions of Ref.~\onlinecite{CammarotaB11}.

%
%

\subsection{\label{sec:MinimaHistograms}Distribution of Local Minima}

To explore this crossover in more detail, we used 
basin-hopping parallel tempering (BHPT) to explore and sample the local minima of pinned BLJ,
using the same reference configuration prepared at $T_0=0.5$.  The aim is to
estimate the density of minima $g_{\rm IS}(V,Q_0)$, 
\red{so the basin-hopping temperatures of the different replicas were selected to promote exploration
of a large variety of minima. We found that for $T_{\rm BH}\gtrsim 25$ the range of minimum energies explored
becomes constant, and replicas with $T_{\rm BH} < 0.5$ do not explore new minima efficiently.
Therefore, we used 12 basin-hopping replicas with $T_{\rm BH}$ spaced
geometrically between $0.5$ and $25.0$, to allow efficient exchange of configurations between replicas.}
 For each replica, 10 basin-hopping runs
of $10^5$ steps were performed, and the results combined to produce a larger
database of minima. The basin-hopping step size was varied dynamically to
ensure that approximately 70\% of steps located a new minimum.

\red{Every basin-hopping replica was initialised at the reference minimum $\mathbf{X}_0$. This choice may bias the sampled distribution of minima towards high-$Q_0$ regions of the landscape, however the inclusion of high-temperature replicas which rapidly decorrelate from the initial minimum should limit this effect.}

\begin{figure}
  \includegraphics[width=0.5\textwidth]{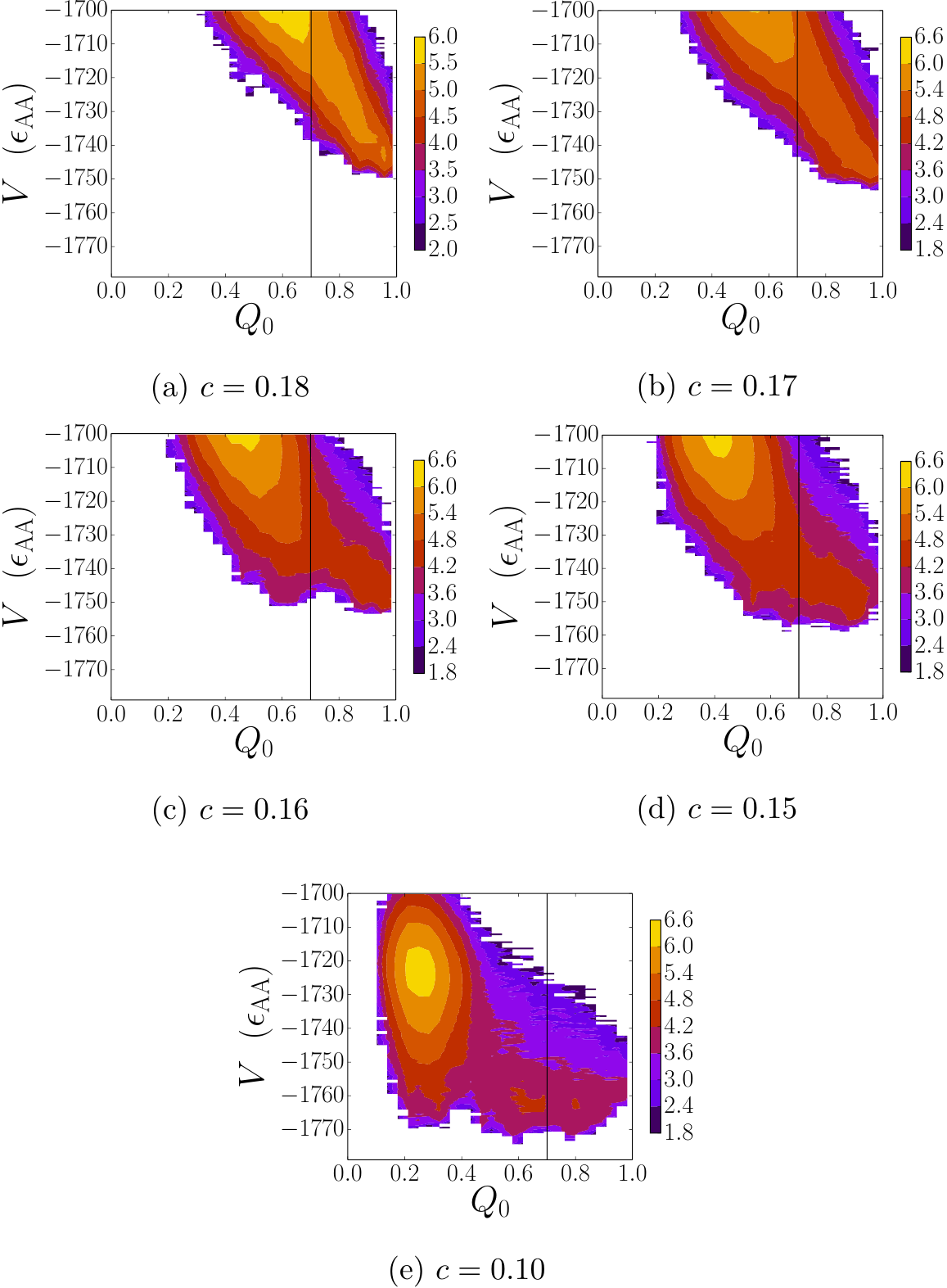}
  \caption{Contour plot of $\log_{10}\rho_{\rm IS}(V,Q_0)$ for databases produced using BHPT. $\rho_{\rm IS}(V,Q_0)$ is proportional to the number of minima in the database that have potential energy $V$ and overlap $Q_0$ with the reference minimum.}
  \label{fig:histograms}
\end{figure}

Fig.~\ref{fig:histograms} shows contour plots of $\log_{10}\rho_{\rm IS}(V,Q_0)$ for several values of $c$.
The essential feature is that for $c\geq 0.17$ then all low-energy minima have $Q_0>0.7$, while for $c\leq 0.16$ there is significant density of states at low energy with $Q_0<0.7$, indicating that multiple metastable states are accessible.  As in Fig.~\ref{fig:returntrajs}, this crossover occurs over a rather narrow range of $c$: the difference between $c=0.16$ and $c=0.17$ corresponds to pinning 3 extra atoms.

Recall from eq.~(\ref{equ:rho-g}) that the sampled distributions $\rho_{\rm
IS}$ in Fig.~\ref{fig:histograms} cannot be interpreted as direct measurements
of $g_{\rm IS}$; these quantities coincide only in the limit of exhaustive
sampling.  However, we expect that regions where $\rho_{\rm IS}(V,Q_0)=0$
probably have $g_{\rm IS}(V,Q_0)\approx 0$, because there are relatively few
minima at low energies and our results are consistent with having sampled
almost all of them.


%

\subsection{\label{sec:PackingProbs}Global Probability of States}

So far, all results have been shown for a single reference configuration.  We have repeated our analysis for two other reference configurations, and five realisations of the disorder for each $\mathbf{X}^\ast$. To study the changing landscape as a function of $c$, we use consistent sequences of disorder realisations where the set of pinned atoms at each $c$ is a subset of the pinned atoms at all higher pinning fractions.

  To summarise the results, we define
\begin{equation} \label{eq:Pref}
  P(Q_0>0.7) = \int^{V_c}_{-\infty}\int_{0.7}^1 \rho_{\rm IS}(V,Q_0) ~\text{d}Q_0\,\text{d}V, 
\end{equation}
where $V_c$ is an energy cutoff that restricts the integral to low energy (accessible) minima.
  
\begin{figure}
  \begin{center}
  \includegraphics[width=0.4\textwidth]{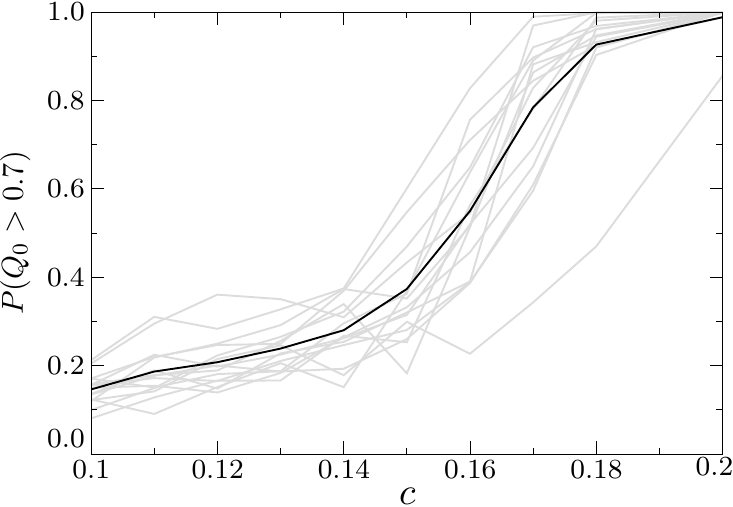}
  \end{center}
  \caption{\label{fig:PofQ}Probability that a low-energy minimum selected at random will be similar to the reference minimum. 15 realisations of the disorder are represented, indicated by grey lines. The black line represents the average value. Three different reference configurations were used, and the 
variation in $P(Q_0>0.7)$ between different disorder realisations with the same reference configuration is similar to the variation between different references.}
\end{figure}

Fig.~\ref{fig:PofQ} shows $P(Q_0>0.7)$ as a function of $c$ for several
sequences of disorder realisations, and the disorder average of this
probability.  For each sequence, one observes a crossover as $c$ is varied.
The position and slope of the crossover vary between realisations, but the
behaviour is similar in all cases, and the average probability also shows a
crossover between $c\approx 0.14$ and $c\approx 0.17$.

%
%
%

\begin{figure}
  \includegraphics[width=0.5\textwidth]{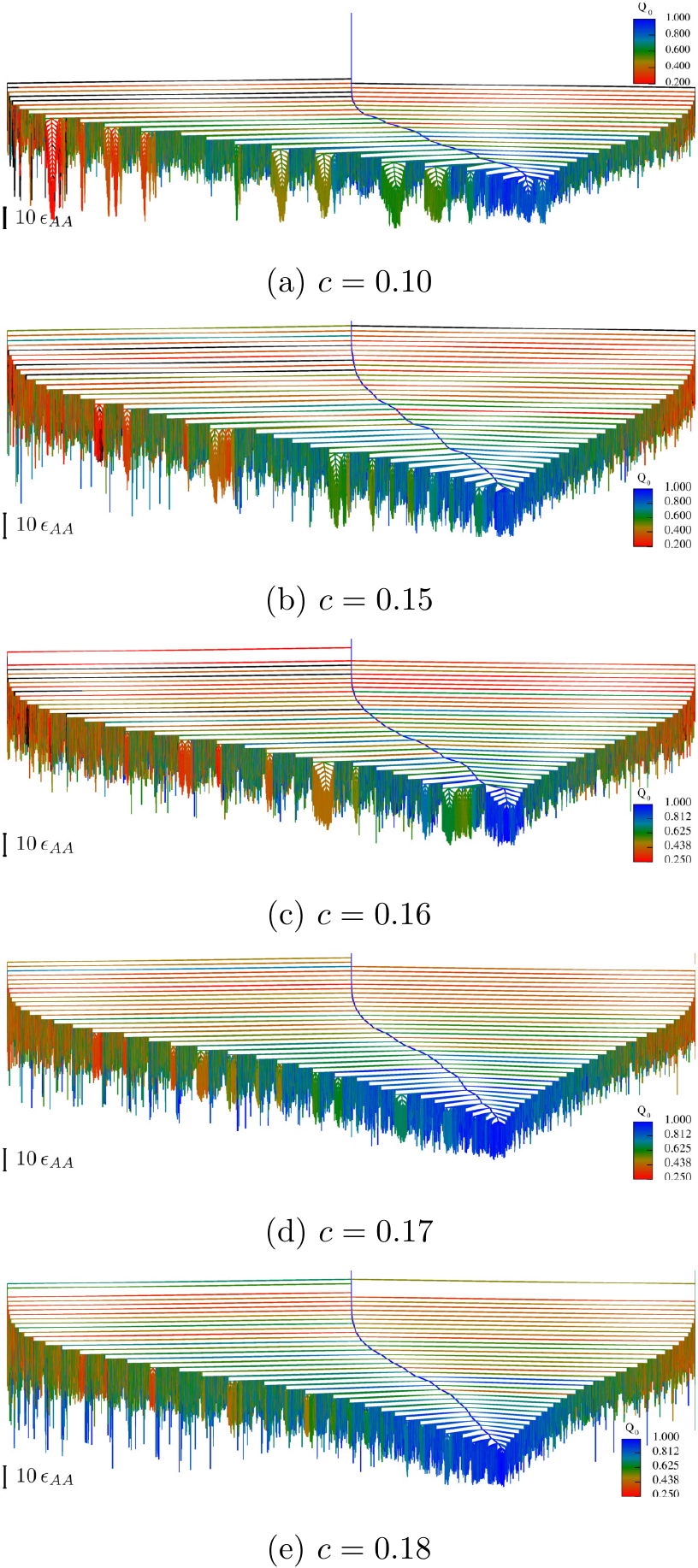}
  \caption{\label{fig:Dgraphs}Disconnectivity graphs for landscapes generated
by discrete path sampling between minima found in BHPT calculations. Minima are
coloured by their value of $Q_0$.}
\end{figure}

\subsection{\label{sec:Landscapes}Transition states}

The results so far 
show that the distribution of local energy minima changes dramatically under random particle pinning. 
However, this analysis considers only the energies of the minima and their overlap $Q_0$ with the reference minimum.  
By projecting the landscape onto these coordinates, one discards a large amount of information, particularly the energies of the transition states (saddle points) that connect the minima. 
There may be multiple funnels/states where minima with similar $Q_0$ are separated by large barriers, which are not apparent from distributions such as $\rho_{\rm IS}(V,Q_0)$.
To resolve this detail, we consider the connectivity of the landscape.



We performed independent transition state sampling for each landscape depicted
in fig.~\ref{fig:histograms}. Initially, 101 local minima were selected from
each database. The reference minimum was always selected, and 100 other
low-energy minima were chosen from a uniform distribution in $Q_0$. We then
used the methods described in Sec.~\ref{sec:TSsearches} to calculate discrete
paths between each pair of minima in the set of 101, a total of 5050 pairs.
Combining the paths yielded a larger database of PEL minima and saddle points
in the region of configuration space spanned by the initial set of 101 minima.
Both low- and high-$Q_0$ regions of space were included.

\subsection{Disconnectivity graphs}

The landscape databases are represented in fig.~\ref{fig:Dgraphs} as
disconnectivity graphs (see Sec.~\ref{sec:DGraphMethods}). Each branch is
coloured according to its overlap $Q_0$ with the reference minimum.
%
The landscape corresponding to $c=0.10$ resembles an unpinned glassy PEL. In particular, it has many local funnels, which we have previously\cite{deSouzaW08} identified with metabasins.\cite{DoliwaH03,DoliwaH03b} There is no dominant lowest-energy region of the PEL, but instead many of the local funnels have comparable energy.  

\begin{figure*}[t]
  \begin{center}
    \includegraphics[width=0.99\textwidth]{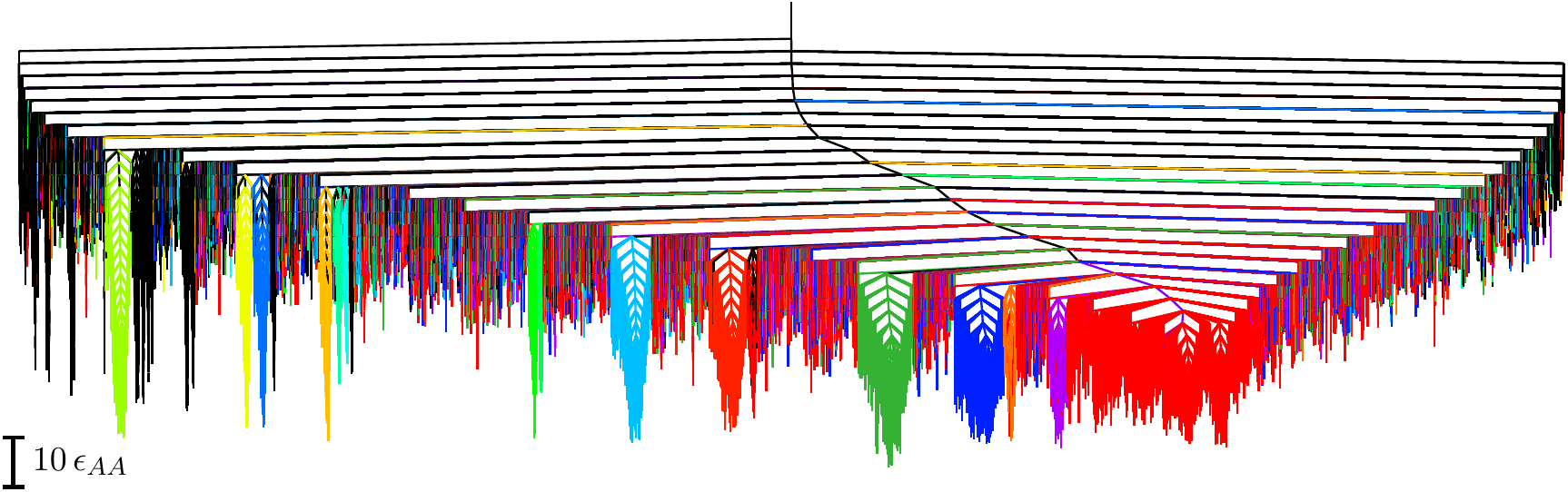}
  \end{center}
  \caption{Disconnectivity graph for the database with $c=0.10$. Minima are assigned to packings using a greedy community detection algorithm, and each packing is coloured differently in the graph. Only packings containing more than 1000 minima are shown, all other minima are coloured black.}
  \label{fig:packingsgraph}
\end{figure*}

At high pinning the PEL has a very different structure. In fig.~\ref{fig:packingsgraph}(e) there is a single low-energy region of the PEL, which contains only minima similar to the reference minimum.  
That is, pinning $18\%$ of the particles has little effect on the landscape funnel that contains the reference state, but it acts to suppress other funnels.
%
%
Single-funnelled landscapes usually correspond to structure-seeking systems,\cite{Wales12} 
which relax efficiently to their global minimum. In fig.~\ref{fig:packingsgraph}(e),
the main funnel contains quite high energy barriers and many minima with
comparable energies, so reproducible relaxation to the global minimum is
unlikely. However, the overall funnel structure means that relaxation to the
region of minima with $Q_0>0.7$ will be fast and irreversible, except at very
high temperatures.
The result of pinning this many particles is that the landscape no longer resembles that of a structural glass former. 
We note, however, that this single-funnelled landscape includes minima with a
wide range of $Q_0$, including low overlaps $Q_0\approx 0.2$.  One should not
imagine that pinning destroys all minima with low overlap.  Rather, one finds
that such minima still exist, but their energies are large, compared to the
reference.



We also recall that Fig.~\ref{fig:histograms} shows a significant difference
between $c=0.16$ and $c=0.17$, where the possibility of minima with low energy
and $Q_0<0.7$ disappears rather suddenly.  The disconnectivity graphs in
Fig.~\ref{fig:Dgraphs} indicate a smoother crossover as $c$ is varied: the
low-overlap funnels that compete with the reference funnel at $c=0.16$ do not
disappear on increasing to $c=0.17$; instead it seems that the energy of these
funnels increases, so that they are no longer competitive with the reference
funnel.  We return to this point in Sec.~\ref{sec:LandscapeEvolution}, below.


As a final comment on Fig.~\ref{fig:Dgraphs}, note that there are a
considerable number of minima with low energy and high $Q_0$, which
nevertheless appear to be separated by large barriers from the reference
funnel.  This is probably due to ``artificial trapping'':\cite{StrodelWW07} it
is likely that there exist low-energy transition states connecting these
minima to the reference funnel, but they have not been sampled in our
transition state search.  However, it is also possible that there are some
large barriers between low-energy minima with high overlap, caused by the
immovable pinned atoms.

\subsection{\label{sec:PackingDGraphs}Packings on the PEL} 

In fig.~\ref{fig:Dgraphs} most local funnels are coloured uniformly, suggesting that the minima within each funnel are structurally similar. 
This similarity is expected in general for landscapes with funnels, particularly in glasses where the funnels are approximately equivalent to dynamical metabasins.\cite{deSouzaW08}



To investigate the relationship between landscape structure and real-space structure, we use $Q$ as a similarity measure.
In particular, we identify sets of minima, such that all minima within each set have a high mutual overlap $Q(\mathbf{X_a},\mathbf{X_b})$, while minima in different communities have low overlap.  Physically, we argue that that these sets correspond to \emph{distinct packings} of the particles, so that different minima within each packing typically differ by small local displacements.  On the other hand, minima in different packings correspond to larger displacements, such as those that happen during structural relaxation of the liquid.



Many  methods exist\cite{Ward63,GirvanN02,SzekelyR05} for detecting highly-connected sets in a graph with edge weights given by a similarity measure, but these typically require evaluation of all edge weights.  Since our databases contain of order $10^5$ minima
we use a greedy algorithm, which is typically much cheaper: 
\begin{enumerate}
  \item The ``parent minimum'', $\mathbf{X}_p$, for the first packing is the reference minimum $\mathbf{X}_0$.
  \item \label{step:checkmin} Compute $Q(\mathbf{X}_p,\mathbf{X}_m)$ for each minimum $\mathbf{X}_m$ not currently assigned to a packing.
  \item \label{step:addmin} If $Q(\mathbf{X}_p,\mathbf{X}_m)>Q^\ast$, add $m$ to the same packing as $p$.
  \item \label{step:newparent} Use the lowest-energy unassigned minimum as $\mathbf{X}_p$ for the next packing.
  \item Iterate steps \ref{step:checkmin}-\ref{step:newparent} until all unassigned minima lie above a predefined energy threshold.
\end{enumerate}


The greedy algorithm does not guarantee that every minimum in a packing is more similar to the parent of that packing than to any other parent. However, it does guarantee that all minima within a packing are structurally close, and all parents are dissimilar to each other. 

Fig.~\ref{fig:packingsgraph} shows the results of the greedy algorithm applied to a database at $c=0.10$. 
Branches are coloured according to the packing to which the corresponding minimum belongs. Most local funnels visible on the landscape correspond to a single packing, verifying that the the order parameter $Q$ can be used to detect whether two minima are in the same funnel.  Of course, two configurations that are dissimilar from the reference minimum (small $Q_0$) might have a high mutual overlap (if they are in the same funnel) or a low mutual overlap, if they are in different funnels.  In this sense the disconnectivity graph contains much more information than histograms such as Fig.~\ref{fig:histograms}, because it reveals the existence of multiple distinct packings/funnels.


However, we note that there are many minima in Fig.~\ref{fig:packingsgraph}
that are identified as members of a packing, but are not members of any funnel.
As in Fig.~\ref{fig:Dgraphs}, this is likely a result of artificial kinetic trapping: 
these minima may be part of a funnel, but the relevant transition state has not
been found. 

\subsection{\label{sec:lowT}Effect of Reference Temperature and System Size}

\begin{figure}
    \begin{center}
      \includegraphics[width=0.5\textwidth]{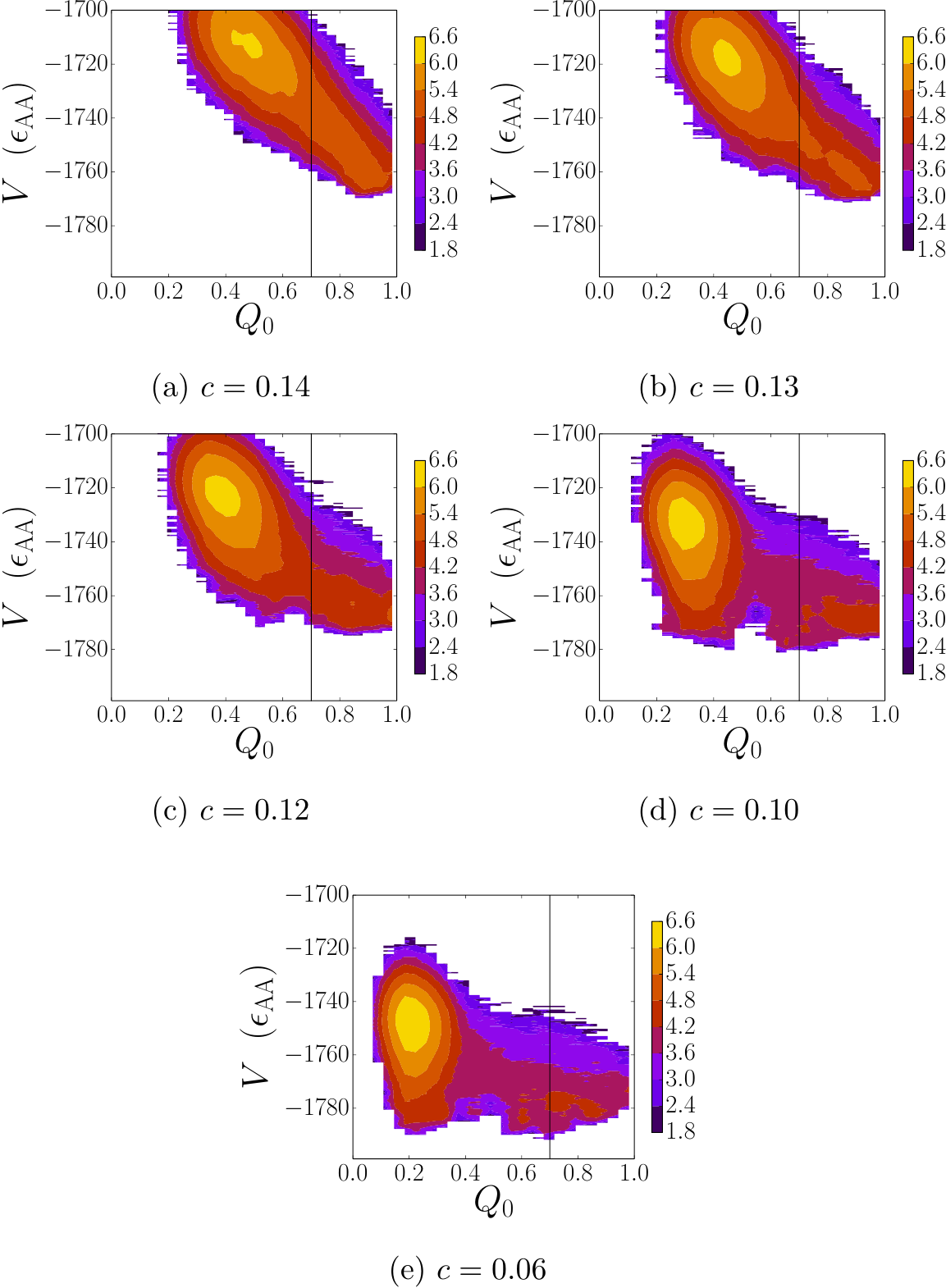}
    \end{center}
  \caption{Density of states $\log_{10}\rho_{\rm IS}(V,Q_0)$ for a sequence of disorder realisations with reference temperature $T_0 = 0.43\,\epsilon_{AA}/k_{\text{B}}$.}
  \label{fig:lowThistograms}
\end{figure}

\begin{figure}
  \begin{center}
  \includegraphics[width=0.4\textwidth]{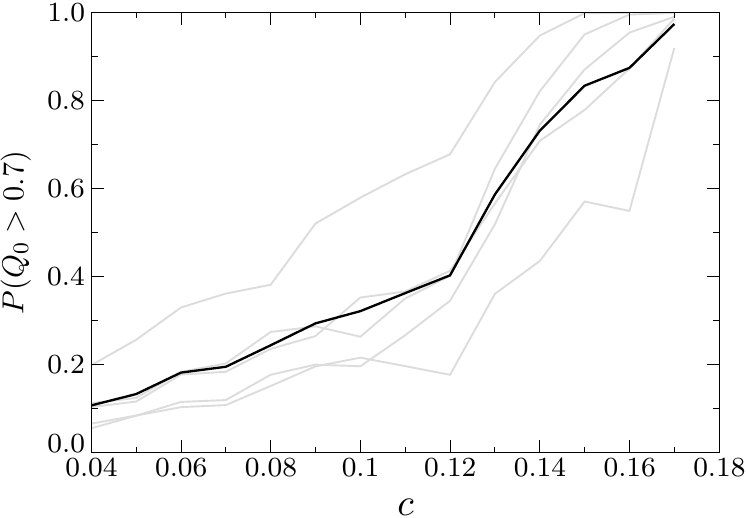}
  \end{center}
  \caption{\label{fig:PofQ-lowT}Probability that a low-energy minimum selected at random will be similar to the reference minimum, with reference temperature $T_0=0.43$. Five different reference configurations have been used. Grey lines indicate calculations from a single reference configuration, the black line is the disorder average.
  }
\end{figure}

So far all results have used reference configurations that are Boltzmann-distributed at $T_0=0.5$. 
Glassy features of the system are expected to become more accentuated on cooling, and mean-field theory predicts that $c^\ast$ decreases with $T_0$.
To test these expectations, we used BHPT to sample energy minima with reference configurations obtained at $T_0=0.43$. 
Results for one reference configuration are shown in Fig.~\ref{fig:lowThistograms}.  Comparing with Fig.~\ref{fig:histograms}, we note that the high-overlap minima are separated in energy from the low-overlap ones for $c>0.13$, while this required a larger pinning fraction $c=0.17$ in Fig.~\ref{fig:histograms}.  
This effect is consistent with the theoretical picture of Ref.~\onlinecite{CammarotaB11}.
Fig.~\ref{fig:PofQ-lowT} summarises the behaviour of five realisations of the disorder.  As noted above, the crossover from low- to high-overlap occurs at a smaller value of $c$, compared with Fig.~\ref{fig:PofQ}: this effect is consistent among different realisations of the disorder.

The dependence of this crossover on $T_0$ and on system size is crucial for the theory of Ref.~\onlinecite{CammarotaB11}, which predicts a thermodynamic 
phase transition as $c$ is increased.  Our results are not sufficient to investigate this prediction  in detail, but Appendix~\ref{sec:SystemSize} 
shows preliminary results for a smaller system ($N=180$ particles).   The smaller systems show a clearer separation between high- and low-overlap, possibly 
because larger systems can support distinct high- and low-overlap regions within the same sample, making the crossover less sharp. 
The methods that we have introduced here provide a natural framework for investigating these questions.

\subsection{Discussion}
\label{sec:xover-discuss}

We summarise the conclusions so far.  From Figs.~\ref{fig:histograms} and
\ref{fig:PofQ}, we see that increasing the number of pinned particles reduces
the number of low-energy minima with significant equilibrium occupation probabilities.  For $T_0=0.5$, there
is a crossover at $c\approx 0.16$ so that for larger $c$, the accessible minima
are mostly in \emph{the same packing} as the reference minimum, as evidenced
by their high overlap values $Q_0$.  From Fig.~\ref{fig:Dgraphs}, one sees
that this crossover is accompanied by a change in the energy landscape, from a
rough (glassy) landscape to a single-funnelled disconnectivity graph.  From
Figs.~\ref{fig:Dgraphs} and~\ref{fig:packingsgraph} one sees that the funnels
observed in the disconnectivity graph are closely related to the
packings that can be identified by analysis of the overlap (recall
Sec.~\ref{sec:funnels-etc}).  Figs.~\ref{fig:lowThistograms} and
\ref{fig:PofQ-lowT} show that lower reference temperatures make these effects
more pronounced, particularly that there is a clearer distinction between
high-overlap and low-overlap minima.

All these results are broadly consistent with the theoretical predictions of
Ref.~\onlinecite{CammarotaB11} and with previous simulation
work.\cite{KobB13,OzawaKIM15} The crossovers that we observe happen at slightly
larger $c$ than predicted by Ref.~\onlinecite{OzawaKIM15}, and the effect of
$T_0$ on the position of the crossover seems to be weaker.  However, the
quantities that we use to characterise this crossover are also quite different,
so one does not expect quantitative agreement.  

\rlj{As noted above, understanding whether
this crossover corresponds to a thermodynamic phase transition would require a more detailed analysis
of different system sizes and temperatures.  From a physical point of view, the first-order character of the RFOT transition
predicts a clear separation between two sets of minima, that correspond to macrostates with high- and low-overlap.  Intermediate values of the overlap should be strongly suppressed
by the interfacial free energy cost associated with coexistence of different macrostates.\cite{KobB13,BerthierJ15,JackG16}  Obtaining a clear separation between the minima with high- and low-overlap
is hindered for $T_0=0.5$ because the relatively large number of pinned particles reduces the possibility of very small  $Q_0$: it appears from  Fig.~\ref{fig:histograms} that the distributions for the (putative) high- and low-$Q$ macrostates are somewhat overlapping, and no clear trough is observed in the probability.
One expects a clearer separation at lower $T_0$, where fewer particles are pinned, but this is not immediately apparent in Fig.~\ref{fig:lowThistograms}.}
\red{The absence of a clear separation between macrostates} \rlj{may be attributed to the relatively small system sizes used here: one expects that the distributions for the two macrostates should become narrower in larger systems, leading to a trough in the probability and (perhaps) to a directly-observable interface between high- and low-overlap states.  Unfortunately, these larger systems are challenging numerically, as they are for other methods~\cite{OzawaKIM15,KobB13}.
For these reasons, we defer this question to later work.}

\rlj{In the remainder of this paper} we discuss several other aspects of the energy landscapes in these pinned systems, particularly the degree of frustration, and the extent to which one can think of metabasins evolving smoothly on the landscape as $c$ is reduced.

\section{Landscape frustration}
\label{sec:frustration}

In this section, we present quantitative descriptions of the change in
landscape organisation as a function of $c$. First, we consider simple
properties, which hint at the change in structure observed in
fig.~\ref{fig:Dgraphs} and then we propose a more sophisticated metric to
quantify this change directly.

\begin{figure}
    \begin{center}
      \includegraphics[width=0.35\textwidth]{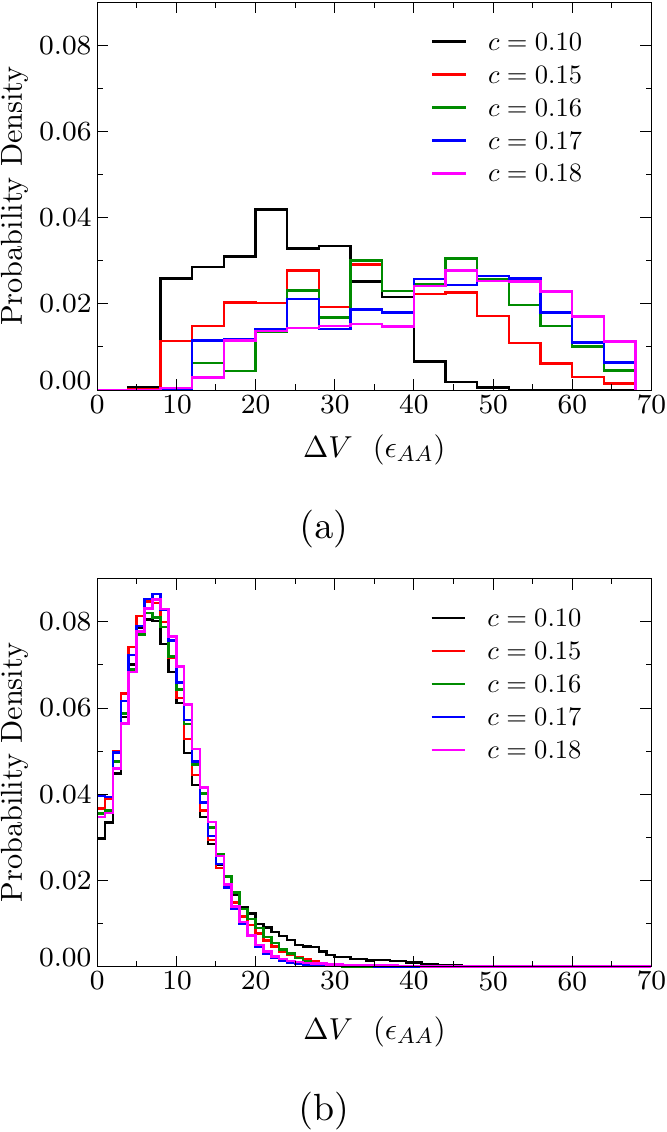}
    \end{center}
  \vspace{-0.5cm}
  \caption{\label{fig:RefBarriers} Histograms of the heights of energy barriers between local minima and the reference minimum. The first panel shows the ``uphill'' barriers, the second panel shows ``downhill'' barriers. The scale is the same for both panels.}
\end{figure}

Fig.~\ref{fig:RefBarriers} shows histograms of the energy barriers between
local minima and the reference minimum for each landscape database. The
barriers are divided into ``uphill'', i.e. barriers to go from the reference
minimum to a particular local minimum, and ``downhill'', from the minimum to
the reference.

The average uphill barrier increases systematically with $c$, because the sides
of the main landscape funnel become steeper, as observed in
fig.~\ref{fig:Dgraphs}. In contrast, the mean downhill barrier is quite
insensitive to $c$, but a long tail of high energy barriers develops as $c$
decreases. This tail corresponds to minima in low-energy funnels that have
higher energy barriers to the reference than minima in the main funnel. 



Fig.~\ref{fig:RefBarriers} and fig.~\ref{fig:Dgraphs} 
both indicate that 
pinned PELs become less frustrated as $c$ increases, meaning that there are fewer low-energy regions of the landscape separated by high barriers. Therefore the \red{simulation time required} to reach high-$Q_0$ minima will be small at high $c$.


Previously, we have characterised PELs using a frustration metric, $\widetilde{f}$, related to the efficiency of \red{locating} the global minimum \red{from a randomly-chosen minimum}.\cite{deSouzaSNFW17} In this case, we are interested in \red{transition rates towards} 
the PEL funnel that contains the reference minimum (\red{which is} not necessarily the global minimum). Therefore a modified frustration metric is used, based on the definition of packings presented in Sec.~\ref{sec:PackingDGraphs}:
\begin{equation}
  \widetilde{f_p} = \sum_{\alpha \notin P_0} \widetilde{p}_\alpha^\text{eq}(T)\left(\frac{V_\alpha^\dag - V_0}{\text{max}\{V_\alpha-V_0,\Delta V\}}\right).
  \label{eq:packingfrustration}
\end{equation}
Here, $P_0$ is the the packing that contains the reference minimum, and $\alpha$
runs over all minima that do not belong to $P_0$. $p_\alpha^{\rm eq}(T)$ is the
equilibrium occupation probability of $\alpha$, calculated within the harmonic
superposition approximation.\cite{Wales93,Wales2003}
$\widetilde{p}_\alpha^\text{eq} = p_\alpha^\text{eq}/(1-\sum_{\beta \in
P_0}p_\beta^\text{eq})$ is the renormalised probability excluding all minima
belonging to $P_0$. $V_0$ and $V_\alpha$ are the energies of the reference
minimum and minimum $\alpha$, respectively. 

\red{$V_\alpha^\dag$ is the energy of the highest transition state on the
minimum-energy pathway connecting $\alpha$ to the reference, 
and $\Delta V=0.1\,\epsilon_{AA}$ is a parameter chosen to avoid divergence of
$\widetilde{f_p}$ in cases where the reference minimum is not the global
minimum.} 

\begin{figure}[t]
    \begin{center}
      \includegraphics[width=0.4\textwidth]{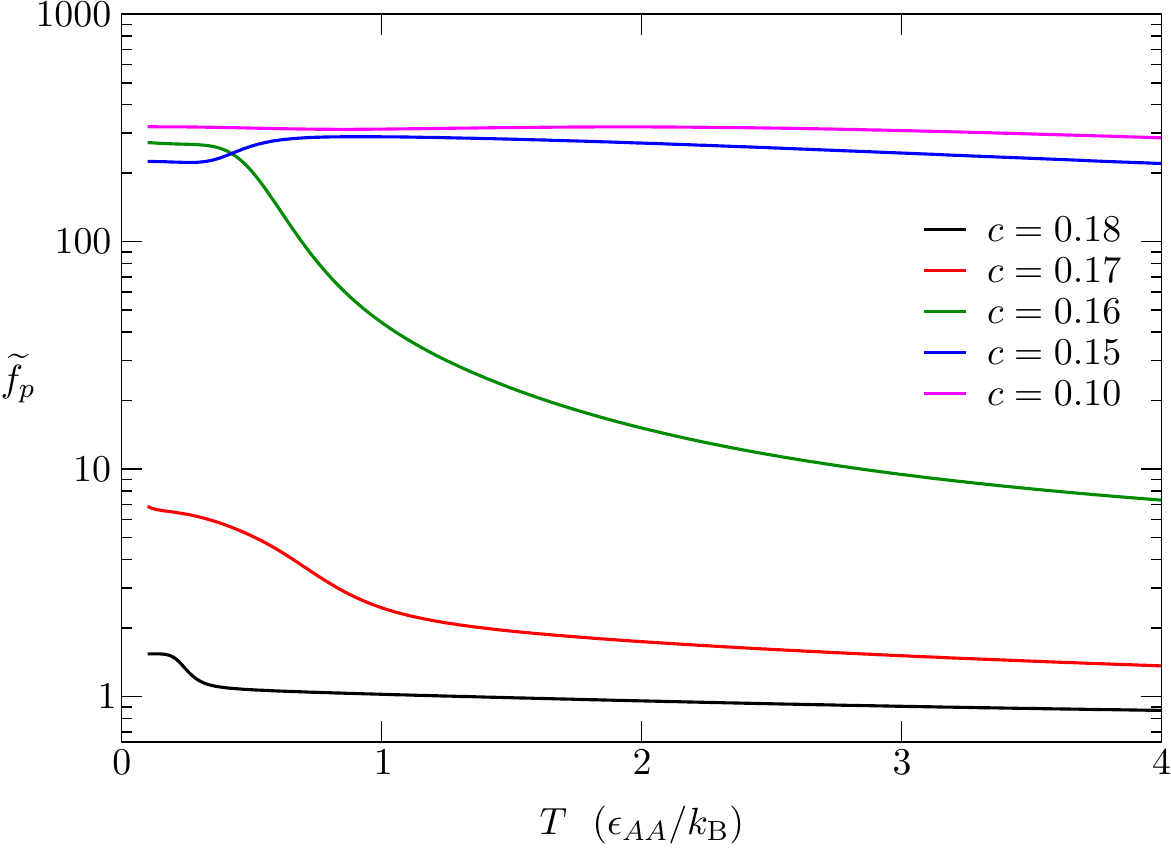}
    \end{center}
    \caption{\label{fig:modfrustration}Modified frustration index $\widetilde{f_p}(T)$ for landscapes with different pinning fraction.}
\end{figure}

$\widetilde{f_p}$ is plotted in fig.~\ref{fig:modfrustration} for the landscapes represented in fig.~\ref{fig:Dgraphs}.
Frustration decreases as a function of $c$ for the entire temperature range
plotted, illustrating \red{a major structural change in the PEL concurrent
 with the pinning transition. Over this range in $c$, the landscape transforms
 from a multifunnelled structure typical of supercooled liquids
into a single-funnelled non-glassy structure.
 This result agrees with and reinforces our qualitative
interpretation of fig.~\ref{fig:Dgraphs}. 
The large range in $\widetilde{f_p}$ emphasises the magnitude of the change.}

At low temperatures, $\widetilde{f_p}(T)$ varies more rapidly because the sum in eq.~\ref{eq:packingfrustration} is dominated by a few large terms. In particular, the frustration of the $c=0.16$ landscape increases dramatically, which may be a peculiarity of this particular disorder realisation, 
because several packings at $c=0.16$ are almost degenerate in energy with the reference minimum.

\section{\label{sec:LandscapeEvolution}Evolution of the PEL as the Pinning Fraction is reduced}


In this section, we examine the effects of random pinning by
following the behaviour of particular minima and packings as $c$ changes.   
For example, we consider the overlap between minima in landscapes that have different $c$, and hence appear in different panels of Fig.~\ref{fig:Dgraphs}.

We perform this analysis by starting from a high-$c$ landscape ($c=c_0=0.18$), 
and unpinning atoms one by one (always in the same sequence), relaxing the PEL minima by energy minimisation after each atom is unpinned. New sets of minima at lower values of $c$ are obtained.

One might imagine a complementary procedure, following the evolution of minima
as $c$ is increased.  This approach would be interesting, but it is
difficult to implement because there is no unique route to obtain a
minimum at $c_0$, given an initial minimum at some lower pinning fraction $c$.
We therefore leave this analysis for future work.  




\subsection{\label{sec:MinimaEvolution}Evolution of the Minima}

We studied the properties of a set of minima during unpinning from $c_0=0.18$. This set included the reference minimum at $c_0$, and the parent minimum of every packing that contained at least 1000 minima.



\begin{figure*}
    \begin{center}
      \includegraphics[width=\textwidth]{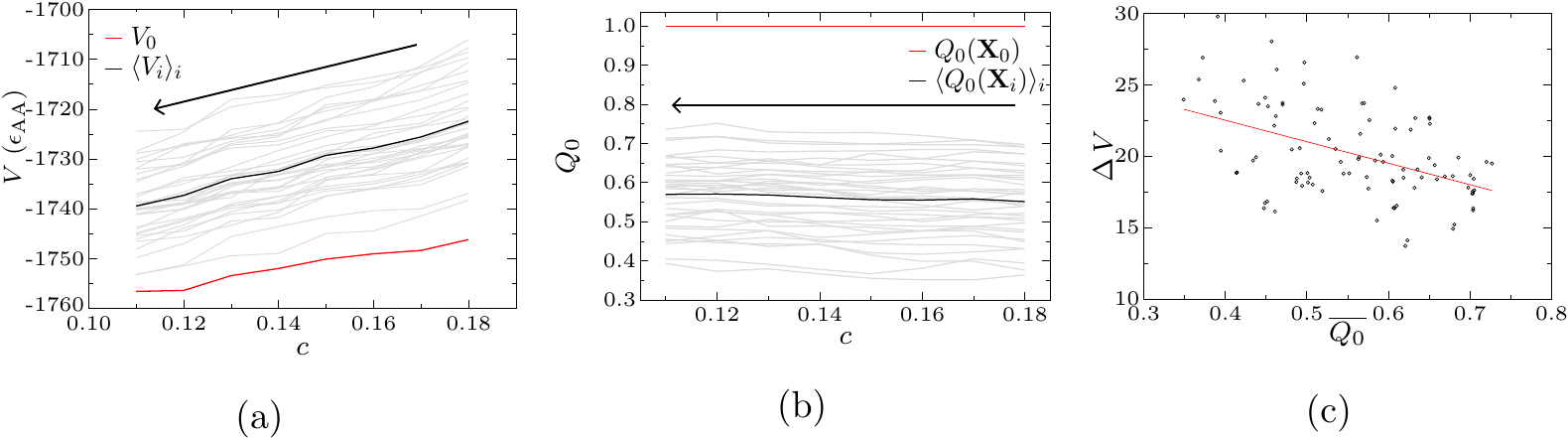}
    \end{center}
   \vspace{-0.5cm}
  \caption{\label{fig:relaxprofiles} Figures showing how $V$ and $Q_0$ evolve for a set of minima with progressive unpinning from a $c=0.18$ landscape (as indicated by the arrows). Each grey line represents the parent minimum of a large packing on the $c=0.18$ landscape. The thick black line represents the average value over the grey lines. The red line represents the reference minimum.
\red{Panel (c) shows how the energy change during unpinning depends on the overlap of the minimum. $\Delta\,V$ is the energy difference between a minimum at $c=0.18$ and the corresponding minimum at $c=0.10$, $\overline{Q_0}$ is the average $Q_0$ of those two minima. The best-fit trend line is shown, which has correlation coefficient -0.463.}}
\end{figure*}

Fig.~\ref{fig:relaxprofiles}(a) shows how the energies of these minima change
during unpinning. The energy of each minimum decreases, because the energy is
reminimised after each particle is unpinned.
For most minima this decrease is substantial: around $20\,\epsilon_{AA}$ on
average. Because the reference minimum decreases by only $10\,\epsilon_{AA}$
over the same interval, this result means that the offset between the reference
minimum and the other funnels decreases during unpinning.
Fig.~\ref{fig:relaxprofiles}(b) shows little change in $Q_0$ during unpinning, indicating that most packings do not undergo significant structural change when pinned atoms are released. 

%
%

\subsection{\label{sec:PELEvolution}Evolution of the Packings}

\begin{figure}[t]
  \centering
  \includegraphics[width=0.5\textwidth]{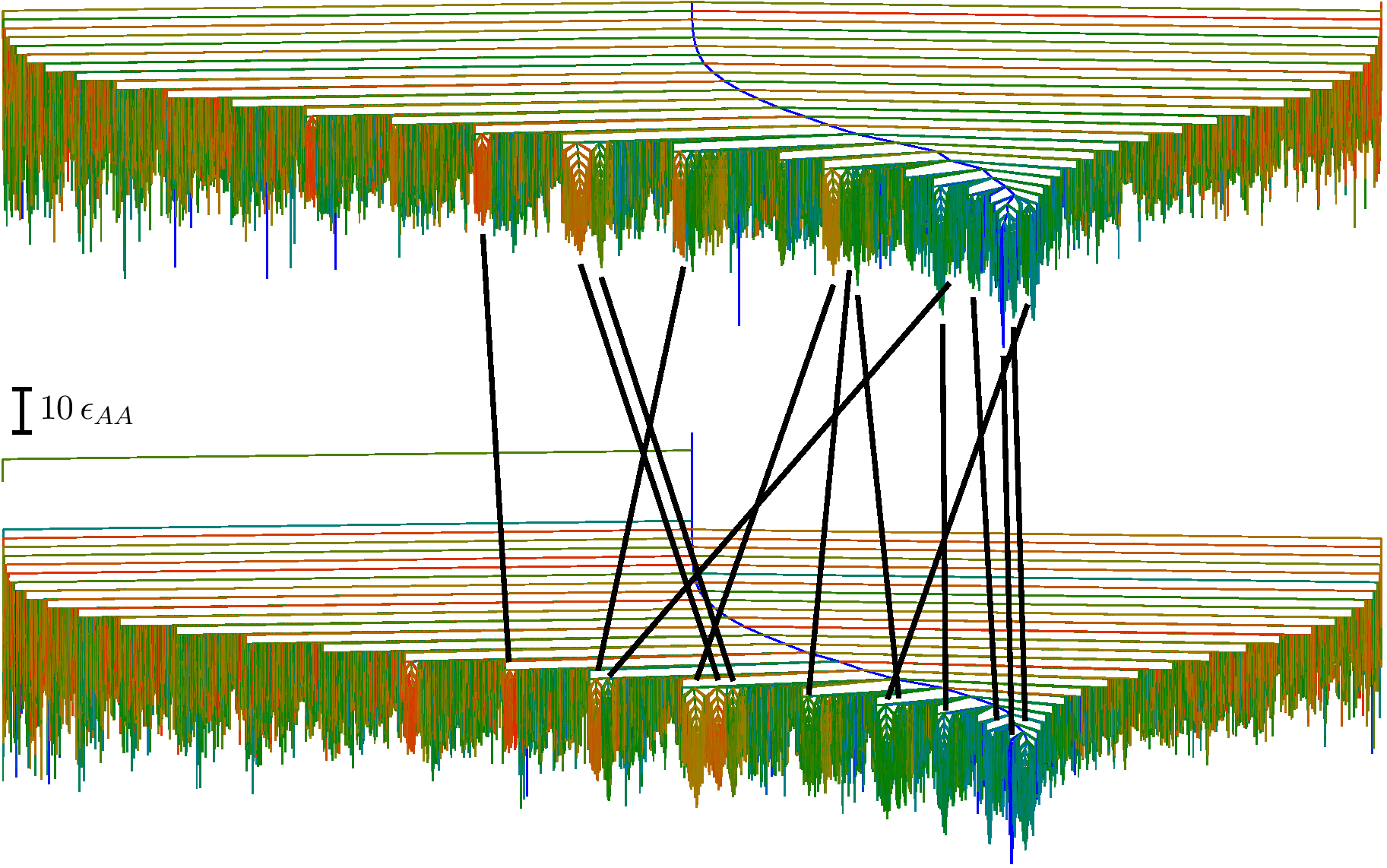}
  \caption{Top panel: correspondence between packings in different disconnectivity graphs. The upper graph represents a landscape with $c=0.17$, the lower graph is at $c=0.16$. Black lines connect some of the packings on the two landscapes that have $\mathcal{Q}>0.7$. Minima are coloured according to their $Q_0$ values, using the same colour scale as before.} 
  \label{fig:packingsevolution}
\end{figure}

We now consider the evolution of \emph{packings} (groups of minima), as $c$ is
decreased.  We took a sample of 101 minima representing all the large packings
at $c=c_0=0.18$, and used the unpinning procedure to obtain the corresponding
minima at $c=0.17$.  For both pinning fractions, we found discrete paths between
every pair of minima, ensuring full connectivity. We also repeated this unpinning
procedure to obtain databases at $c=0.16$ and $c=0.15$.
Fig.~\ref{fig:packingsevolution} shows the disconnectivity graphs for $c=0.17$
and $c=0.16$.

We emphasise that the disconnectivity graph obtained by unpinning in fig.~\ref{fig:packingsevolution} is not at all equivalent to the graphs shown in fig.~\ref{fig:Dgraphs}, even if the value of $c$ is the same.  
The set of minima used to initialise the path-sampling calculation in fig.~\ref{fig:Dgraphs} were obtained by BHPT sampling at the same value of $c$ as the transition states. In contrast, the minima used for path-sampling in fig.~\ref{fig:packingsevolution} were obtained by relaxing minima that were sampled by BHPT at a higher value of $c$. 
\red{Therefore the disconnectivity graphs in fig.~\ref{fig:packingsevolution} may be thought of as the subset of the $c=0.17$ and $c=0.16$ landscapes that is directly related to minima that also exist on the $c=0.18$ landscape, whereas fig.~\ref{fig:Dgraphs} represents a sample of the entire PEL at each value of $c$.}

\red{Note the differences between the two figures: in the $c=0.16$ panel of fig.~\ref{fig:Dgraphs} there are several low-$Q_0$ packings with energies within 1-2\,$\epsilon_{\rm AA}$ of the reference structure, but the gap in the $c=0.16$ panel of fig.~\ref{fig:packingsevolution} is significantly larger.} 
\rlj{Our methods do not allow us to follow minima as $c$ increases, but the natural conclusion here is that the packings that exist (for $c=0.16$) with low energy and low $Q_0$ are somehow ``projected out'' as the number of pinned particles is increased, which explains why they have no counterparts in the high-$c$ landscape.  This is consistent with the theory of random pinning.\cite{CammarotaB11}}



\begin{figure}
    \begin{center}
      \includegraphics[width=0.4\textwidth]{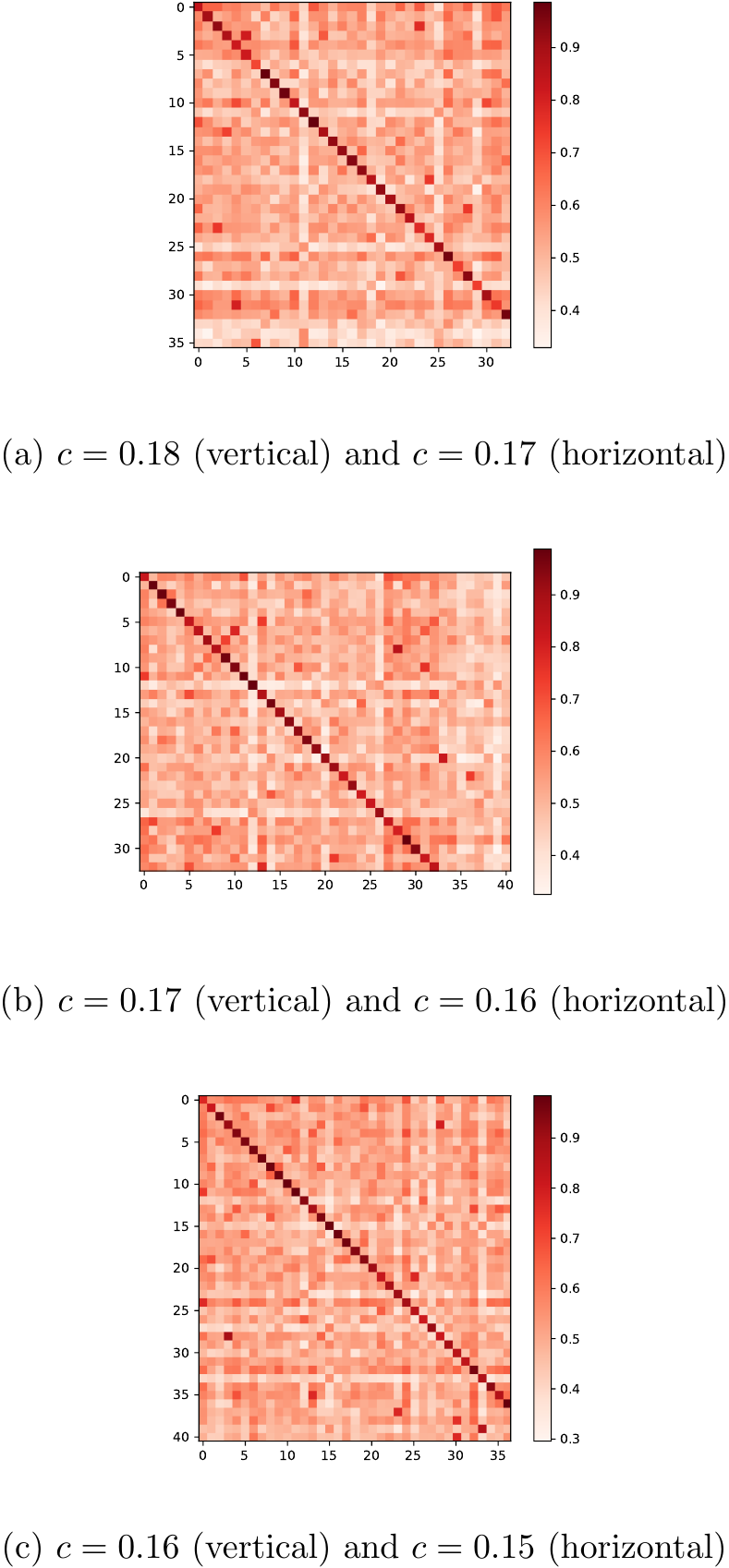}
    \end{center} 
\caption{\label{fig:overlapmatrices} Heat maps of matrices representing the average overlap between low-lying minima in different packings for two landscapes with different $c$. 
Packings have been ordered to maximise overlap along the main diagonal.}
\end{figure}

We used the packing-detection algorithm of Sec~\ref{sec:PackingDGraphs} to identify packings in both landscapes, restricting to packings that 
contain at least 1000 minima.  We estimated the mutual overlap between pairs of packings: for two packings $A,B$ we define
$$
{\cal Q}(A,B) = \frac{1}{{\cal N}_A{\cal N}_B} \sum_{X\in A} \sum_{Y\in B} Q(X,Y)
$$
where the sums run over all minima within each packing (the number of minima in packing $A$ is ${\cal N}_A$, etc).
In practice, we estimate $\cal Q$ by selecting 10 minima at random from each packing.
The sum in the overlap calculation includes all atoms that are unpinned in the lower-$c$ configuration.

Fig.~\ref{fig:overlapmatrices} shows ${\cal Q}(A,B)$ for different pairs of
landscapes.  The labelling of the packings is arbitrary; they have been ordered
using the Hungarian algorithm\cite{Kuhn55,Munkres57} to maximise the overlap
along the main diagonal of each panel. Most packings at the lower value of $c$
have high overlap with exactly one ``parent packing'' from the higher $c$.
Some of the correspondences between parent and daughter packings are shown in
Fig.~\ref{fig:packingsevolution}.  As in Fig.~\ref{fig:relaxprofiles}, one sees
that packings retain their identities as $c$ is reduced, but \red{fig.~\ref{fig:packingsevolution}
also shows that the energy gap between the reference and low-$Q_0$ minima decreases slightly
as $c$ is reduced. This observation indicates a weak negative relationship between $Q_0$ and 
the energy decrease during unpinning, which is illustrated by fig.~\ref{fig:relaxprofiles}(c).} 

As well as pairs of packings that have a clear parent-daughter relationship,
there are several other scenarios that can (and do) occur.  First, there may be
packings on the low-$c$ landscape that have no apparent parent on the high-$c$
landscape.  These features correspond to columns in Fig.~\ref{fig:overlapmatrices} in
which no large values appear.  In this case, unpinning leads to new packings
that were not present at higher $c$, consistent with an increasing
value of $S_{\rm MS}$ as $c$ is reduced.  Second, there may be packings on
the high-$c$ landscape that have no clear daughter on the low-$c$ landscape --
these correspond to rows in Fig.~\ref{fig:overlapmatrices} with no large
values.  In this case, unpinning some atoms has presumably led to a significant
rearrangement in the structure of the system -- the original packing may have
been stabilised by one of the pinned atoms, and is destroyed by unpinning.
Third, there may be daughter packings (at low $c$) with more than one parent;
and there may be parent packings (at high $c$) with more than one daughter.
These correspond to splitting or merging of packings as $c$ is reduced.
Fourth, we sometimes observe two parent and two daughter packings, such that
both parents have high overlap with both daughters: in
Fig.~\ref{fig:overlapmatrices} one then sees an off-diagonal element with a
large value of $\cal Q$, together with a large value of $\cal Q$ in the
corresponding transposed element.  This scenario indicates pairs of packings that
are structurally similar, but not similar enough to be identified as a single
packing by our packing-detection algorithm.

To end this section, recall that the original theoretical picture of random pinning~\cite{CammarotaB11} is that 
$S_{\rm MS}$ is reduced as $c$ is increased, leading to an RPGT when there is only one packing
with appreciable occupation probability. 
\rlj{Here we have considered the unpinning process (decreasing $c$), which limits our ability to draw conclusions about the behaviour when $c$ increases.
In particular, since the packings shown in Fig.~\ref{fig:overlapmatrices} are obtained by successive unpinning, one tends not to sample low-$c$ packings 
that lack any ``parent'' (in the higher-$c$ landscape).} \red{To the extent that this limitation may be ignored,} \rlj{our results follow the qualitative behavior predicted by mean-field theory,\cite{CammarotaB11} 
although situations in which some packings have multiple parents or multiple daughters are not expected in mean-field models.}


\section{\label{sec:Conclusions}Conclusions}

We have analysed the energy landscape of a randomly-pinned glassy fluid, using
the overlap $Q$ as an order parameter.  As the pinning fraction $c$ is
increased, the energy landscape crosses over from a typical glassy structure
with many funnels, into a single-funnelled structure in which all
thermally-accessible minima have high overlap with the reference minimum.  These  
observations match the situation anticipated by Cammarota and Biroli,\cite{CammarotaB11}
although the data presented here cannot resolve whether this crossover
corresponds to the predicted thermodynamic phase transition.  We showed
that the overlap, which is the natural order parameter within mean-field
theories of the glass transition,\cite{FranzP97,MezardP99} can be used to
define distinct packings of the particles, and that these packings can be
identified with funnels on the energy landscape. \red{We propose that packings
represent physically relevant metastable states, such that the associated entropy
$S_{\rm MS}$ should vanish at the RPGT. Future numerical work could test this
hypothesis.}

We quantified the change in landscape structure by calculating a frustration
metric, which indicates that \red{thermodynamic and dynamic bias towards the
reference structure is greater at high $c$ than low $c$.}
In addition, we
introduced methods for tracking packings (and individual minima), as $c$ is
reduced.  The results indicate a complex phenomenology, where packings
typically seem to retain their identity as $c$ changes, but they can
also split and merge.

The methodology presented here offers a new route for investigating the effects
of random pinning in amorphous systems, providing a link between mean-field
theories and the potential energy landscape.

\begin{acknowledgments}
\noindent 

This article may be downloaded for personal use only. Any other use requires prior permission of the author and AIP Publishing. This article appeared in {\it J. Chem. Phys.} {\bf 149}, 114503 (2018) and may be found at https://doi.org/10.1063/1.5042140.

Research data supporting this article is freely available for download at https://doi.org/10.17863/CAM.30987.

This work was supported by the University of Cambridge, through a CHSS studentship to S.P.N., and by the EPSRC.

\end{acknowledgments}
\input{jnames}

\appendix
\section{\label{sec:SystemSize}Effect of System Size}

\begin{figure}
    \begin{center}
      \includegraphics[width=0.5\textwidth]{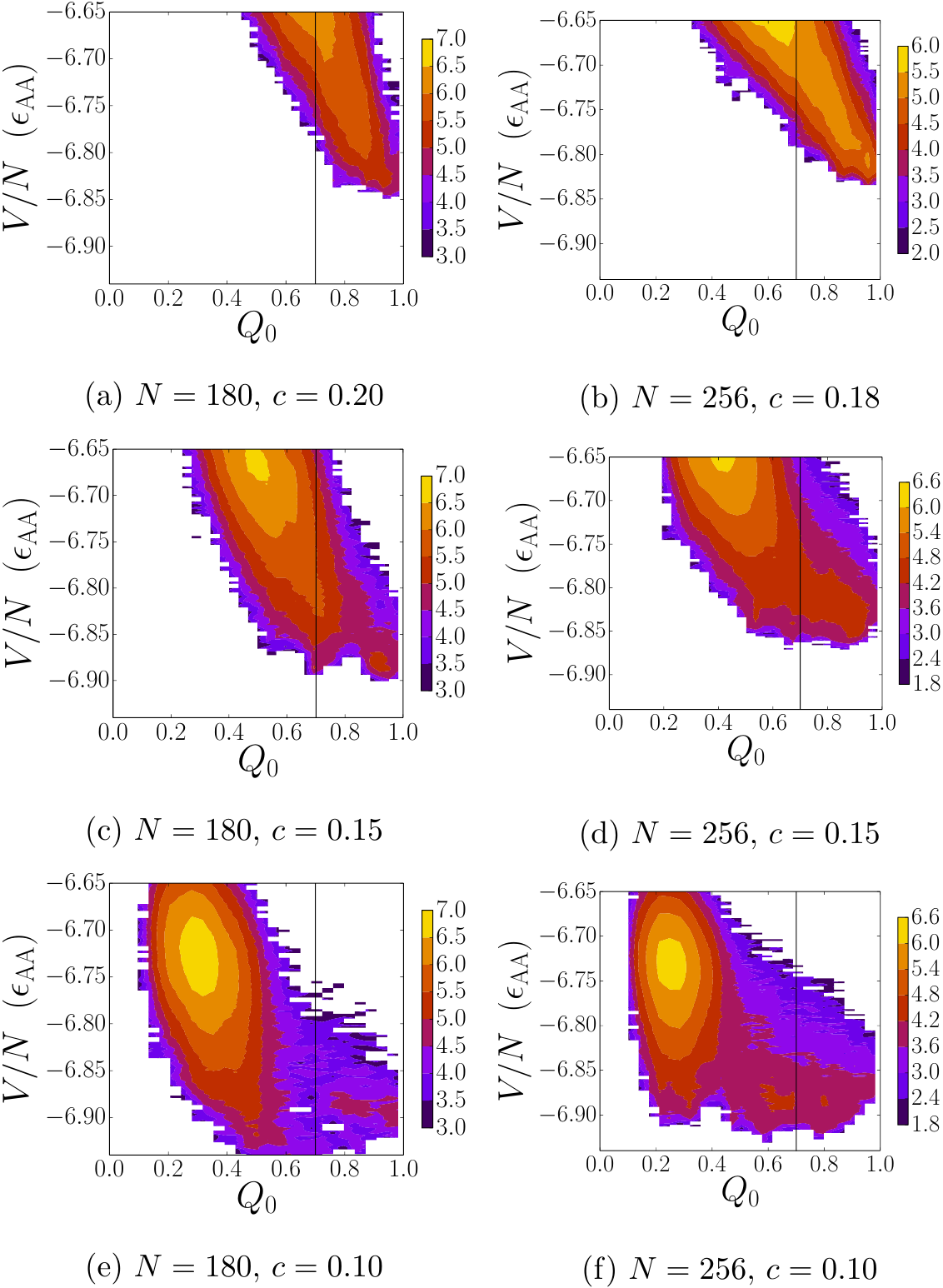}
    \end{center}
  \caption{Histograms comparing $\rho_{\rm IS}(V,Q_0)$ between different system sizes. All graphs have the same vertical scale.}
  \label{fig:size_comparison}
\end{figure}


If the RPGT is an equilibrium phase transition, $P(Q_0>0.7)$ should be
discontinuous in the thermodynamic limit, because the number of distinct
metastable states would change suddenly at $c^\ast$. Finite-size effects suppress
this discontinuity,\cite{Yeomans1992,Goldenfeld1992} but can in principle be
removed by system size scaling analysis. The gradient of $P(Q_0>0.7)$ at the
crossover should extrapolate to infinity in the infinite system-size limit if
the landscape transformation corresponds to a thermodynamic event.
\rlj{Also, in large systems, one expects to see a bimodal distribution $\rho_{\rm IS}(V,Q_0)$, with two populations of 
minima (high- and low-overlap), that are separated by a deep trough in the probability. This trough represents the 
interfacial-free-energy cost to nucleate a configuration with low-overlap, within a high-overlap system.}

As an initial step towards this scaling analysis, 
fig.~\ref{fig:size_comparison} presents $\rho_{\rm IS}(V,Q_0)$ for a smaller BLJ
simulation cell containing 180 atoms (144 A-type and 36 B-type), and comparable
plots for the 256-atom landscapes. No qualitative differences are observed
between the two systems, which may indicate that the landscape properties we
probe are broadly independent of system size. The density of
minima with $Q_0$ values intermediate between the high- and low-overlap states
(i.e.~$Q_0\approx 0.7$) is slightly smaller in the smaller system. \rlj{This result may
suggest that the larger system is better able to support distinct high- and
low-$Q_0$ regions within the same configuration, which tends to ``smooth out'' the sharp transition predicted  in mean-field theory
(or push it to lower temperature).}

\bibliography{pinning.bib}

\end{document}

%% file: jnames.tex

\newcommand\acie{Angew. Chem. Int. Ed.\xspace}
\newcommand\aciee{Angew. Chem. Int. Ed. Engl.\xspace}
\newcommand\ac{Acta. Crystallogr.\xspace}
\newcommand\acis{Adv. Colloid Interface Sci.\xspace}
\newcommand\acp{Adv. Chem. Phys.\xspace}
\newcommand\acr{Acc. Chem. Res.\xspace}
\newcommand\aicej{Amer. Inst. Chem. Eng. J.\xspace}
\newcommand\aip{Adv. Phys.\xspace}
\newcommand\ajp{Am. J. Phys.\xspace}
\newcommand\am{Adv. Mater.\xspace}
\newcommand\ams{Ann. Math. Stat.\xspace}
\newcommand\apb{Appl. Phys. B\xspace}
\newcommand\aps{Adv. Polym. Sci.\xspace}
\newcommand\aqc{Adv. Quant. Chem.\xspace}
\newcommand\arbbc{Ann. Rev. Biophys. Biophys. Chem.\xspace}
\newcommand\armb{Ann. Rev. Microbiol.\xspace}
\newcommand\arpc{Ann. Rev. Phys. Chem.\xspace}
\newcommand\bbpc{Ber. Bunsenges. Phys. Chem.\xspace}
\newcommand\bc{Biochemistry\xspace}
\newcommand\bi{Bioinf.\xspace}
\newcommand\bmk{Biometrika\xspace}
\newcommand\bmm{Biomacromolecules\xspace}
\newcommand\bp{Biopolymers\xspace}
\newcommand\bpc{Biophys. Chem.\xspace}
\newcommand\bpj{Biophys. J.\xspace}
\newcommand\cb{Chem. Brit.\xspace}
\newcommand\cccc{Coll. Czech. Chem. Comm.\xspace}
\newcommand\cms{Comp. Mat. Sci.\xspace}
\newcommand\cocis{Curr. Op. Colloid Interface Sci.\xspace}
\newcommand\cop{Comm. Phys.\xspace}
\newcommand\cosb{Curr. Op. Struct. Biol.\xspace}
\newcommand\cpc{Comp. Phys. Comm.\xspace}
\newcommand\cphyc{Chem. Phys. Chem.\xspace}
\newcommand\cpl{Chem. Phys. Lett.\xspace}
\newcommand\cpr{Comp. Phys. Rep.\xspace}
\newcommand\cps{Colloid Polym. Sci.\xspace}
\newcommand\crev{Chem. Rev.\xspace}
\newcommand\ea{Electrochim. Acta\xspace}
\newcommand\el{Europhys. Lett.\xspace}
\newcommand\epjb{Eur. Phys. J. B\xspace}
\newcommand\epjd{Eur. Phys. J. D\xspace}
\newcommand\epje{Eur. Phys. J. E\xspace}
\newcommand\fd{Faraday Discuss.\xspace}
\newcommand\febsl{FEBS Lett.\xspace}
\newcommand\fpe{Fluid Phase Equilib.\xspace}
\newcommand\ic{Inorg. Chem.\xspace}
\newcommand\ijmpc{Int. J. Mod. Phys. C\xspace}
\newcommand\ijqc{Int. J. Quant. Chem.\xspace}
\newcommand\irc{Int. Rev. Cytol.\xspace}
\newcommand\irpc{Int. Rev. Phys. Chem.\xspace}
\newcommand\jasa{J. Am. Stat. Assoc.\xspace}
\newcommand\jcis{J. Colloid Interface Sci.\xspace}
\newcommand\jcsft{J. Chem. Soc., Faraday Trans.\xspace}
\newcommand\jacers{J. Am. Ceram. Soc.\xspace}
\newcommand\jacs{J. Am. Chem. Soc.\xspace}
\newcommand\jas{J. Atmos. Sci.\xspace}
\newcommand\jbc{J. Biol. Chem.\xspace}
\newcommand\jcc{J. Comp. Chem.\xspace}
\newcommand\jchp{J. Chim. Phys.\xspace}
\newcommand\jce{J. Chem. Ed.\xspace}
\newcommand\jcop{J. Comp. Phys.\xspace}
\newcommand\jcscc{J. Chem. Soc., Chem. Commun.\xspace}
\newcommand\jetp{J. Exp. Theor. Phys. (Russia)\xspace}
\newcommand\jmb{J. Mol. Biol.\xspace}
\newcommand\jmsc{J. Mat. Sci.\xspace}
\newcommand\jmgm{J. Mol. Graph. Mod.\xspace}
\newcommand\jmsp{J. Mol. Spec.\xspace}
\newcommand\jmst{J. Mol. Struct.\xspace}
\newcommand\jncs{J. Non-Cryst. Solids\xspace}
\newcommand\jnet{J. Non-Equilib. Thermodyn.\xspace}
\newcommand\jpa{J. Phys. A\xspace}
\newcommand\jpb{J. Phys. B\xspace}
\newcommand\jpc{J. Phys. Chem.\xspace}
\newcommand\jpca{J. Phys. Chem. A\xspace}
\newcommand\jpcb{J. Phys. Chem. B\xspace}
\newcommand\jpcc{J. Phys. Chem. C\xspace}
\newcommand\jpcm{J. Phys. Cond. Mat.\xspace}
\newcommand\jpcs{J. Phys. Chem. Solids\xspace}
\newcommand\jpf{J. Phys. France\xspace}
\newcommand\jphc{J. Phys. C\xspace}
\newcommand\jpsj{J. Phys. Soc. Japan\xspace}
\newcommand\jr{J. Rheol.\xspace}
\newcommand\jrssb{J. R. Stat. Soc. B\xspace}
\newcommand\jsp{J. Stat. Phys.\xspace}
\newcommand\jvsta{J. Vac. Sci. Technol. A\xspace}
\newcommand\lang{Langmuir\xspace}
\newcommand\mg{Math. Gazette\xspace}
\newcommand\mm{Macromolecules\xspace}
\newcommand\molp{Mol. Phys.\xspace}
\newcommand\ms{Mol. Simul.\xspace}
\newcommand\msec{Mat. Sci. Eng. C\xspace}
\newcommand\mts{Macromol. Theory Simul.\xspace}
\newcommand\nc{Nuovo Cimento\xspace}
\newcommand\nl{Nano Lett.\xspace}
\newcommand\njp{New J.~Phys.\xspace}
\newcommand\nrg{Nature Rev.: Genetics\xspace}
\newcommand\nmat{Nature Mater.\xspace}
\newcommand\nps{Nature Phys. Sci.\xspace}
\newcommand\nsb{Nature Struct. Biol.\xspace}
\newcommand\nt{Nucl. Technol.\xspace}
\newcommand\osid{Open Sys. Inf. Dyn.\xspace}
\newcommand\pa{Physica A\xspace}
\newcommand\paa{Proc. Amsterdam Acad.\xspace}
\newcommand\pac{Pure. Appl. Chem.\xspace}
\newcommand\pca{Physica\xspace}
\newcommand\pccp{Phys. Chem. Chem. Phys.\xspace}
\newcommand\pcps{Prog. Colloid. Polym. Sci.\xspace}
\newcommand\pb{Phys. Biol.\xspace}
\newcommand\pbmb{Prog. Biophys. Mol. Biol.\xspace}
\newcommand\pd{Physica D\xspace}
\newcommand\phys{Physics\xspace}
\newcommand\pkm{Phys. Kond. Materie\xspace}
\newcommand\pla{Phys. Lett. A\xspace}
\newcommand\plos{PLOS Biol.\xspace}
\newcommand\pmaga{Philos. Mag. A\xspace}
\newcommand\pmagb{Philos. Mag. B\xspace}
\newcommand\pnas{Proc. Natl. Acad. Sci. USA\xspace}
\newcommand\pnasu{Proc. Natl. Acad. Sci. USA\xspace}
\newcommand\ppmsj{Proc. Phys.-Math. Soc. Japan\xspace}
\newcommand\pr{Phys. Rev.\xspace}
\newcommand\prep{Phys. Reports\xspace}
\newcommand\prsa{Proc. R. Soc. A\xspace}
\newcommand\pe{Prot. Eng.\xspace}
\newcommand\ps{Prot. Sci.\xspace}
\newcommand\psfb{Proteins: Struct., Func. and Bioinf.\xspace}
\newcommand\psfg{Proteins: Struct., Func. and Gen.\xspace}
\newcommand\pt{Phys. Today\xspace}
\newcommand\ptps{Prog. Theor. Phys. Supp.\xspace}
\newcommand\ptrsla{Philos. Trans. Roy. Soc. Lond. A\xspace}
\newcommand\rpp{Rep. Prog. Phys.\xspace}
\newcommand\sci{Science\xspace}
\newcommand\sa{Sci. Amer.\xspace}
\newcommand\ssci{Surf. Sci.\xspace}
\newcommand\ssp{Solid State Phys.\xspace}
\newcommand\str{Struct.\xspace}
\newcommand\tbt{Trends Biotech.\xspace}
\newcommand\tbs{Trends Biochem. Sci.\xspace}
\newcommand\tca{Theor. Chim. Acta\xspace}
\newcommand\tfs{Trans. Faraday. Soc.\xspace}
\newcommand\vir{Virology\xspace}
\newcommand\zpb{Z. Phys. B.\xspace}
\newcommand\zpc{Z. Phys. Chem.\xspace}
\newcommand\zpd{Z. Phys. D\xspace}
\newcommand\jctc{J. Chem. Theor. Comput.\xspace}

%% file: pinning_flat.bbl
\begin{thebibliography}{10}

\bibitem{BerthierB11}
Berthier, L. and Biroli, G., \emph{Rev. Mod. Phys.\xspace}, \textbf{83}, 587
  (2011)

\bibitem{LubchenkoW07}
Lubchenko, V. and Wolynes, P.G., \emph{\arpc}, \textbf{58}, 235 (2007)

\bibitem{ChandlerG10}
Chandler, D. and Garrahan, J.P., \emph{Annual Review of Physical Chemistry},
  \textbf{61}, 191 (2010)

\bibitem{gotzes88}
G\"otze, W. and Sj\"ogren, L., \emph{J. Phys. C.}, \textbf{21}, 3407 (1988)

\bibitem{adamg65}
Adam, G. and Gibbs, J.H., \emph{\jcp}, \textbf{43}, 139 (1965)

\bibitem{CammarotaB11}
Cammarota, C. and Biroli, G., \emph{\pnas}, \textbf{109}, 8850 (2012)

\bibitem{ScheidlerKBP02}
Scheidler, P., Kob, W., Binder, K. and Parisi, G., \emph{Philosophical Magazine
  B}, \textbf{82}, 283 (2002)

\bibitem{KimMS11}
Kim, K., Miyazaki, K. and Saito, S., \emph{Journal of Physics: Condensed
  Matter}, \textbf{23}, 234123 (2011)

\bibitem{BerthierK12}
Berthier, L. and Kob, W., \emph{Physical Review E - Statistical, Nonlinear, and
  Soft Matter Physics}, \textbf{85}, 011102 (2012)

\bibitem{KobB13}
Kob, W. and Berthier, L., \emph{Phys. Rev. Lett.}, \textbf{110}, 245702 (2013)

\bibitem{SzamelF13}
Szamel, G. and Flenner, E., \emph{Europhys. Lett.}, \textbf{101}, 66005 (2013)

\bibitem{FullertonJ14}
Fullerton, C.J. and Jack, R.L., \emph{Physical Review Letters}, \textbf{112},
  255701 (2014)

\bibitem{OzawaKIM15}
Ozawa, M., Kob, W., Ikeda, A. and Miyazaki, K., \emph{\pnas}, \textbf{112},
  6914 (2015)

\bibitem{CammarotaS16}
Cammarota, C. and Seoane, B., \emph{Phys. Rev. B}, \textbf{94}, 180201 (2016)

\bibitem{ChakrabartyDKD16}
Chakrabarty, S., Das, R., Karmakar, S. and Dasgupta, C., \emph{J. Chem. Phys.},
  \textbf{145}, 034507 (2016)

\bibitem{Wales2003}
Wales, D.J., \emph{Energy Landscapes},  (Cambridge University Press,
  Cambridge,~2003)

\bibitem{Goldstein69}
Goldstein, M., \emph{\jcp}, \textbf{51}, 3728 (1969)

\bibitem{StillingerW82}
Stillinger, F.H. and Weber, T.A., \emph{Phys. Rev. A}, \textbf{25}, 978 (1982)

\bibitem{StillingerW84}
Stillinger, F.H. and Weber, T.A., \emph{Science}, \textbf{225}, 983 (1984)

\bibitem{SastryDS98}
Sastry, S., Debenedetti, P.G. and Stillinger, F.H., \emph{Nature},
  \textbf{393}, 554 (1998)

\bibitem{SchroderSDG00}
Schroder, T.B., Sastry, S., Dyre, J.C. and Glotzer, S.C., \emph{J. Chem.
  Phys.}, \textbf{112}, 9834 (2000)

\bibitem{DoliwaH03}
Doliwa, B. and Heuer, A., \emph{Phys. Rev. E}, \textbf{67}, 031506 (2003)

\bibitem{DoliwaH03b}
Doliwa, B. and Heuer, A., \emph{Phys. Rev. E}, \textbf{67}, 030501(R) (2003)

\bibitem{Heuer08}
Heuer, A., \emph{J. Phys. Cond. Mat.}, \textbf{20}, 373101 (2008)

\bibitem{deSouzaW08}
de~Souza, V.K. and Wales, D.J., \emph{J. Chem. Phys.}, \textbf{129}, 164507
  (2008)

\bibitem{NiblettDSW16}
Niblett, S.P., de~Souza, V.K., Stevenson, J.D. and Wales, D.J., \emph{J. Chem.
  Phys.}, \textbf{145}, 024505 (2016)

\bibitem{lis87}
Li, Z. and Scheraga, H.A., \emph{Proc. Natl. Acad. Sci. USA}, \textbf{84}, 6611
  (1987)

\bibitem{walesd97a}
Wales, D.J. and Doye, J.P.K., \emph{J. Phys. Chem. A}, \textbf{101}, 5111
  (1997)

\bibitem{Wales02}
Wales, D.J., \emph{Mol. Phys.}, \textbf{100}, 3285 (2002)

\bibitem{Wales04}
Wales, D.J., \emph{Mol. Phys.}, \textbf{102}, 891 (2004)

\bibitem{FranzP97}
Franz, S. and Parisi, G., \emph{\prl}, \textbf{79}, 2486 (1997)

\bibitem{MezardP99}
M\'ezard, M. and Parisi, G., \emph{Phys. Rev. Lett.}, \textbf{82}, 747 (1999)

\bibitem{deSouzaSNFW17}
De~Souza, V.K., Stevenson, J.D., Niblett, S.P., Farrell, J.D. and Wales, D.J.,
  \emph{J. Chem. Phys.}, \textbf{146}, 124103 (2017)

\bibitem{KobA95a}
Kob, W. and Andersen, H., \emph{Phys. Rev. E}, \textbf{51}, 4626 (1995)

\bibitem{ChakrabartyKD15}
Chakrabarty, S., Karmakar, S. and Dasgupta, C., \emph{Scientific Reports},
  \textbf{5}, 12577 (2015)

\bibitem{deSouzaW05}
de~Souza, V.K. and Wales, D.J., \emph{J. Chem. Phys.}, \textbf{123}, 134504
  (2005)

\bibitem{deSouzaW06}
de~Souza, V.K. and Wales, D.J., \emph{Phys. Rev. B.}, \textbf{74}, 134202
  (2006)

\bibitem{deSouzaW06b}
de~Souza, V.K. and Wales, D.J., \emph{Phys. Rev. Lett.}, \textbf{96}, 057802
  (2006)

\bibitem{deSouzaW09}
de~Souza, V.K. and Wales, D.J., \emph{J. Chem. Phys.}, \textbf{130}, 194508
  (2009)

\bibitem{NiblettBWD17}
Niblett, S.P., Biedermann, M., Wales, D.J. and de~Souza, V.K., \emph{\jcp},
  \textbf{147}, 152726 (2017)

\bibitem{StoddardF73}
Stoddard, S.D. and Ford, J., \emph{Phys. Rev. A}, \textbf{8}, 1504 (1973)

\bibitem{ThirumalaiMK89}
Thirumalai, D., Mountain, R.D. and Kirkpatrick, T.R., \emph{Phys. Rev. A},
  \textbf{39}, 3563 (1989)

\bibitem{ThirumalaiM93}
Thirumalai, D. and Mountain, R.D., \emph{Phys. Rev. E}, \textbf{47}, 479 (1993)

\bibitem{Nocedal80}
Nocedal, J., \emph{Math. Comput.}, \textbf{35}, 773 (1980)

\bibitem{LiuN89}
Liu, D. and Nocedal, J., \emph{Math. Prog.}, \textbf{45}, 503 (1989)

\bibitem{BiroliBCGV08}
Biroli, G., Bouchaud, J.P., Cavagna, A., Grigera, T.S. and Verrocchio, P.,
  \emph{Nature Physics}, \textbf{4}, 771 (2008)

\bibitem{Berthier13}
Berthier, L., \emph{Phys. Rev. E}, \textbf{88}, 022313 (2013)

\bibitem{DasCK17}
Das, R., Chakrabarty, S. and Karmakar, S., \emph{Soft Matter}, \textbf{13},
  6929 (2017)

\bibitem{BiroliC17}
Biroli, G. and Cammarota, C., \emph{Phys. Rev. X}, \textbf{7}, 011011 (2017)

\bibitem{KobC14}
Kob, W. and Coslovich, D., \emph{Phys. Rev. E}, \textbf{90}, 052305 (2014)

\bibitem{JonkerV87}
Jonker, R. and Volgenant, A., \emph{Computing}, \textbf{38}, 325 (1987)

\bibitem{WalesC12}
Wales, D.J. and Carr, J.M., \emph{Journal of Chemical Theory and Computation},
  \textbf{8}, 5020 (2012)

\bibitem{BerthierBBKMR07}
Berthier, L., Biroli, G., Bouchaud, J.P., Kob, W., Miyazaki, K. and Reichman,
  D.R., \emph{The Journal of Chemical Physics}, \textbf{126}, 184503 (2007)

\bibitem{BerthierJ15}
Berthier, L. and Jack, R.L., \emph{Phys. Rev. Lett.}, \textbf{114}, 205701
  (2015)

\bibitem{SciortinoKT99}
Sciortino, F., Kob, W. and Tartaglia, P., \emph{\prl}, \textbf{83}, 3214 (1999)

\bibitem{StrodelLWW10}
Strodel, B., Lee, J.W.L., Whittleston, C.S. and Wales, D.J., \emph{Journal of
  the American Chemical Society}, \textbf{132}, 13300 (2010)

\bibitem{TrygubenkoW04}
Trygubenko, S.A. and Wales, D.J., \emph{J. Chem. Phys.}, \textbf{120}, 2082
  (2004)

\bibitem{SheppardTH08}
Sheppard, D., Terrell, R. and Henkelman, G., \emph{\jcp}, \textbf{128}, 134106
  (2008)

\bibitem{HenkelmanUJ00}
Henkelman, G., Uberuaga, B.P. and J\'onsson, H., \emph{J. Chem. Phys.},
  \textbf{113}, 9901 (2000)

\bibitem{HenkelmanJ00}
Henkelman, G. and J\'onsson, H., \emph{J. Chem. Phys.}, \textbf{113}, 9978
  (2000)

\bibitem{munrow99}
Munro, L.J. and Wales, D.J., \emph{Phys. Rev. B}, \textbf{59}, 3969 (1999)

\bibitem{ZengXH14}
Zeng, Y., Xiao, P. and Henkelman, G., \emph{\jcp}, \textbf{140}, 044115 (2014)

\bibitem{Dijkstra59}
Dijkstra, E.W., \emph{Numerische Mathematik}, \textbf{1}, 269 (1959)

\bibitem{CarrTW05}
Carr, J.M., Trygubenko, S.A. and Wales, D.J., \emph{J. Chem. Phys.},
  \textbf{122}, 234903 (2005)

\bibitem{BeckerK97}
Becker, O.M. and Karplus, M., \emph{J. Chem. Phys.}, \textbf{106}, 1495 (1997)

\bibitem{BiroliK01}
Biroli, G. and Kurchan, J., \emph{\pre}, \textbf{64}, 016101 (2001)

\bibitem{Cavagna09}
Cavagna, A., \emph{Phys. Rep.\xspace}, \textbf{476}, 51 (2009)

\bibitem{BerthierC14}
Berthier, L. and Coslovich, D., \emph{\pnas}, \textbf{111}, 11668 (2014)

\bibitem{KurchanL11}
Kurchan, J. and Levine, D., \emph{\jpa}, \textbf{44}, 035001 (2011)

\bibitem{Wales12}
Wales, D.J., \emph{Phil. Trans. Roy. Soc. A}, \textbf{370}, 2877 (2012)

\bibitem{StrodelWW07}
Strodel, B., Whittleston, C.S. and Wales, D.J., \emph{\jacs}, \textbf{129},
  16005 (2007)

\bibitem{Ward63}
Ward, J.H., \emph{Journal of the American Statistical Association},
  \textbf{58}, 236 (1963)

\bibitem{GirvanN02}
Girvan, M. and Newman, M.E.J., \emph{\pnas}, \textbf{99}, 7821 (2002)

\bibitem{SzekelyR05}
Szekely, G.J. and Rizzo, M.L., \emph{Journal of Classification}, \textbf{22},
  151 (2005)

\bibitem{JackG16}
Jack, R.L. and Garrahan, J.P., \emph{\prl}, \textbf{116}, 055702 (2016)

\bibitem{Wales93}
Wales, D.J., \emph{Molecular Physics}, \textbf{78}, 151 (1993)

\bibitem{Kuhn55}
Kuhn, H.W., \emph{Naval Research Logistics Quarterly}, \textbf{2}, 83 (1955)

\bibitem{Munkres57}
Munkres, J., \emph{Journal of the Society for Industrial and Applied
  Mathematics}, \textbf{5}, 32 (1957)

\bibitem{Yeomans1992}
Yeomans, J.M., \emph{Statistical Mechanics of Phase Transitions},  (Clarendon
  Press, Oxford,~1992)

\bibitem{Goldenfeld1992}
Goldenfeld, N., \emph{Lectures on Phase Transitions and the Renormalization
  Group},  (Addison-Wesley, Reading, MA, ~1992)

\end{thebibliography}
